\documentclass[aps,pra,twocolumn,showpacs,superscriptaddress]{revtex4} 
\usepackage{graphicx}
\usepackage{epstopdf}
\usepackage{amsmath}
\usepackage{amssymb}
\usepackage{epstopdf}
\usepackage{amsfonts}
\usepackage{dcolumn}
\usepackage{bigints}
\usepackage{xcolor}
\usepackage{hyperref}
\usepackage[normalem]{ulem}

\usepackage{dirtytalk}

\begin{document}
\title{Auxiliary-state facilitated phase synchronization phenomena in isolated spin systems}
\author{Xylo Molenda}
\address{Homer L. Dodge Department of Physics and Astronomy,
  The University of Oklahoma,
  440 W. Brooks Street,
  Norman,
Oklahoma 73019, USA}
\address{Center for Quantum Research and Technology,
  The University of Oklahoma,
  440 W. Brooks Street,
  Norman,
Oklahoma 73019, USA}
\author{S. Zhong}
\address{Homer L. Dodge Department of Physics and Astronomy,
  The University of Oklahoma,
  440 W. Brooks Street,
  Norman,
Oklahoma 73019, USA}
\address{Center for Quantum Research and Technology,
  The University of Oklahoma,
  440 W. Brooks Street,
  Norman,
Oklahoma 73019, USA}
\author{B. Viswanathan}
\address{Homer L. Dodge Department of Physics and Astronomy,
  The University of Oklahoma,
  440 W. Brooks Street,
  Norman,
Oklahoma 73019, USA}
\address{Center for Quantum Research and Technology,
  The University of Oklahoma,
  440 W. Brooks Street,
  Norman,
Oklahoma 73019, USA}
\author{Xingli Li}
\address{Department of Physics, The Chinese University of Hong Kong, Hong Kong SAR, China}
\author{Y. Yan}
\address{Department of Physics, The Chinese University of Hong Kong, Hong Kong SAR, China}
\author{A. M. Marino}
\address{Homer L. Dodge Department of Physics and Astronomy,
  The University of Oklahoma,
  440 W. Brooks Street,
  Norman,
Oklahoma 73019, USA}
\address{Center for Quantum Research and Technology,
  The University of Oklahoma,
  440 W. Brooks Street,
  Norman,
Oklahoma 73019, USA}
\address{Quantum Information Science Section, Computational Sciences and Engineering Division,
Oak Ridge National Laboratory, Oak Ridge, Tennessee 37831, USA}\thanks{This manuscript has been authored in part by UT-Battelle, LLC, under contract DE-AC05-00OR22725 with the US Department of Energy (DOE). The publisher acknowledges the US government license to provide public access under the DOE Public Access Plan (http://energy.gov/downloads/doe-public-access-plan).}
\author{D. Blume}
\address{Homer L. Dodge Department of Physics and Astronomy,
  The University of Oklahoma,
  440 W. Brooks Street,
  Norman,
Oklahoma 73019, USA}
\address{Center for Quantum Research and Technology,
  The University of Oklahoma,
  440 W. Brooks Street,
  Norman,
Oklahoma 73019, USA}
\date{\today}

\begin{abstract}
Extending classical synchronization to the quantum domain is of great interest both from the
fundamental physics point of view and with a view toward quantum technology applications. This
work characterizes phase synchronization of an effective spin-1 system, which is realized by coupling three quantum states with infinite lifetime to auxiliary excited states that have a finite lifetime. Integrating out the excited states, the effective 
spin-1 model features coherent and incoherent effective  couplings.
Our key findings are: (i) Phase synchronization can be controlled by adjusting the phases of the 
couplings to the excited states.
(ii) Unlike in the paradigmatic spin-1 system studied in the literature, where the dissipative couplings describe decay into the limit cycle state, the effective spin-1 model investigated in this work is governed by a competition between dissipative 
decay into and out of the limit cycle state, with the dissipative decay out of the limit cycle state playing a critical role.
(iii) We identify a parameter regime where phase synchronization of the effective spin-1 system
is---in the absence of coherent effective couplings---governed entirely by the effective dissipators.
The effective spin-1 model is benchmarked through comparisons with master equation calculations for the full
Hilbert space. Physical insights are gained through analytical perturbation theory calculations.
Our findings, which are expected to hold for a broad class of energy level and coupling schemes,
are demonstrated using hyperfine states of $^{87}$Rb.
\end{abstract}
\maketitle

\section{Introduction}

Synchronization of spin systems, which are characterized by a finite number of energy levels, has attracted a great deal of attention over the past few years~\cite{goychuk2006,zhirov2008,xu2014,shankar2017,laskar2020,roulet2018a,roulet2018,koppenhofer2020,parralopez2020,tan2022,zhang2023,tindall2020,kehrer2024, waechtler2024,schmolke2022,schmolke2024,nadolny2023,tao2025}. The finite number of states  distinguishes spin systems from systems that live in  infinite-dimensional Hilbert spaces, such as the quantum mechanical harmonic oscillator~\cite{lee2013,walter2014,dutta2019,mok2020,arosh2021,li2025,xia2025,shen2023,sudler2024,liu2025}.
The finite-dimensional Hilbert space implies that spin systems have no direct classical analog. A classical analog can, however, be established through a two-step procedure in which one first takes the spin $S$ to infinity and then maps to classical equations of motion by replacing operators in  Heisenberg's equations of motion by c-numbers. Since the physics may change significantly as $S$ changes from, e.g., $1/2$, $1$, or $3/2$ to infinity, the  quantum-classical correspondence established through the outlined approach may provide limited insights into the behaviors of small spin systems. 
Because of this,  studies of spin systems with small $S$ offer unique insights into quantum synchronization  phenomena. Changing the number of energy levels by one, e.g., can introduce qualitative changes.  An example is the so-called synchronization blockade~\cite{tan2022}, which has been reported to exist for synchronizable spin systems, with unique behavior for half-integer and integer spins.

  Throughout, we are interested in the quasi-stationary dynamics, and in particular synchronization, of a spin-1 system in the presence of a coherent external drive and dissipative processes.  
Quite generically, synchronization requires the existence of a limit cycle, stabilized through the competition of  linear and non-linear dissipative terms, in the absence of the external drive~\cite{pikovsky2001, vanderpol1926,buca2022}.
In spin systems, the non-linear terms emerge {\em{de facto}} as a consequence of the finite number of energy levels (namely, $2S+1$). 
The paradigmatic driven spin-1 system has been studied extensively~\cite{laskar2020,roulet2018a,roulet2018,tan2022,koppenhofer2019} and is generally considered to be the simplest  model 
that exhibits the synchronization blockade for specific $\gamma_g/\gamma_d$ ratios, where $\gamma_g$ and $\gamma_d$ characterize the rates of the dissipative decays into the limit cycle state~\cite{tan2022,roulet2018a}. The paradigmatic spin-1 model provides a building block for spin chains  as well as spin-oscillator hybrid systems~\cite{tindall2020,kehrer2024,jaseem2020}. 

This work provides a thorough theoretical analysis of an effective spin-1 model, in which coherent and incoherent effective couplings
are realized by coupling to excited states, which have a finite lifetime and sufficiently low population so that 
their existence can be accounted for through effective couplings within the low-energy space (i.e., the Hilbert space of the spin-1 system of interest). Our model is directly relevant to experimental efforts.  
Experimentally, synchronization of a spin-1 system, e.g., was reported in pioneering work by Laskar {\em{et al.}}~\cite{laskar2020}, where effective couplings within the ground state manifold were realized by coupling to auxiliary states. This work shows that the inclusion of the auxiliary states can give rise to intriguing new 
phenomena, distinct from the plethora of synchronization phenomena already discussed in the literature 
for the ``ideal spin-1 system,'' which features coherent couplings and  dissipative decays into the limit cycle state~\cite{laskar2020,roulet2018,tan2022,koppenhofer2019, roulet2018a}.
The inclusion of the excited auxiliary energy states and their subsequent elimination introduces new dissipative
decay paths that lead, as we show in this work, to distinct phase synchronization characteristics.
Our work does not only have  implications for synchronization studies but is, more generally, relevant to  dissipation engineering efforts that utilize atomic, condensed matter, and hybrid  platforms~\cite{harrington2022,verstraete2009,begoc2025,li2025a}.

The remainder of this work is structured as follows.
Section~\ref{secII} introduces the ideal spin-1 model as well as the effective spin-1 system considered in this work. Section~\ref{secIII} presents our result. Finally, Sec.~\ref{sec_outlook} concludes. 
Technical details are relegated to several appendices.

\section{Theoretical background}
\label{secII}

\subsection{Review of the ``ideal spin-$S$ system''}
We are interested in the steady-state dynamics of a spin-$S$ system in the presence of a coherent external drive and dissipative couplings.  
Synchronization of an isolated spin-$S$ system with hermitian Hamiltonian $\hat{H}_0$ is
most commonly described in the rotating frame,
where  $\hat{H}_{0}$  
is equal to $\Delta \hat{S}_z$ and 
the external coherent drive $\hat{H}_D$ is given by $i (\Omega e^{-i\phi_{S}} \hat{S}_- -\Omega e^{i\phi_{S}}\hat{S}_+)/2$~\cite{footnote1}.  In writing $\hat{H}_D$, counter-rotating terms have been dropped. As usual, the ladder operators $\hat{S}_{\pm}$ are equal to $\hat{S}_x \pm i \hat{S}_y$ and $\hat{S}_x$, $\hat{S}_y$, and $\hat{S}_z$ denote the three components of the spin operator $\hat{\vec{S}}$  
(we use the convention that the spin operators are dimensionless). 
Correspondingly, the detuning  
$\Delta$ and coupling $\Omega$, which are assumed to be real, have units of energy.
For $S=1$, the system Hamiltonian $\hat{H}_S$,
$\hat{H}_{S}=\hat{H}_{0}+\hat{H}_D$, has the matrix form
\begin{equation}
\label{eq_spin1_H}
    \underline{H}_{S}
    =\left(
    \begin{array}{ccc}
    \Delta
&
-i \frac{\Omega \exp\left(i\phi_{S}\right)}{\sqrt{2}}
&
0
\\
i \frac{\Omega \exp \left( -i\phi_{S}\right)}{\sqrt{2}}
&
0
&
-i \frac{\Omega \exp\left(i\phi_{S}\right)}{\sqrt{2}}
\\
0
&
i \frac{\Omega \exp\left(-i\phi_{S}\right)}{\sqrt{2}}
&
-\Delta
    \end{array}
    \right),
\end{equation}
where the eigenstates of $\hat{S}_z$ are chosen as the basis states, using the ordering  $|M_S=+1\rangle$, $|M_S=0\rangle$, and $|M_S=-1\rangle$.
The dissipative couplings are typically captured by two dissipators $\hat{\cal{D}}$ with rates $\gamma_g$ and $\gamma_d$~\cite{roulet2018,tan2022},
namely 
$\gamma_g \hat{\cal{D}}[\hat{{L}}_g](\hat{\rho})$
and 
$\gamma_d \hat{\cal{D}}[\hat{{L}}_d](\hat{\rho})$, where $\hat{\rho}$ denotes the density matrix (operator) and
\begin{eqnarray}
    \hat{\cal{D}}[\hat{{L}}](\hat{\rho})=
    \hat{{L}}
    \hat{\rho} \hat{{L}}^{\dagger}-\frac{1}{2} \hat{{L}}^{\dagger} \hat{{L}} \hat{\rho}
    -\frac{1}{2} \hat{\rho}\hat{{L}}^{\dagger} \hat{{L}}. 
\end{eqnarray}
For the spin-1 system, 
the Lindbladians $\hat{{L}}_g$ and $\hat{{L}}_d$ read $\hat{{L}}_g=-2^{-1/2}\hat{S}_{+}\hat{S}_z$ and $\hat{{L}}_d=2^{-1/2}\hat{S}_{-}\hat{S}_z$~\cite{roulet2018,tan2022} (the minus sign in $\hat{L}_g$ does not impact the physics but is included for mathematical reasons): the former can be interpreted as a dissipative coupling from state $|M_S=-1\rangle$ to
state $|M_S=0\rangle$ (gain) while the latter can be interpreted as a dissipative coupling from state $|M_S=+1\rangle$ to
state $|M_S=0\rangle$ (damping).   The zero-temperature master equation, expressed in the rotating frame,  reads
\begin{eqnarray}
\label{eq_me_standard}
\dot{\hat{\rho}}=
-i
[\hat{H}_{S},\hat{\rho}]+
\gamma_g \hat{\cal{D}}[\hat{{{L}}}_g](\hat{\rho})+
\gamma_d \hat{\cal{D}}[\hat{{{L}}}_d](\hat{\rho}).
\end{eqnarray}
In the absence of the external drive ($\Omega=0$), the stationary state $\hat{\rho}_{\text{ss}}$ is equal to $|M_S=0\rangle \langle M_S=0|$, regardless of the initial state. 
 Intuitively, the existence of this limit cycle state can be understood as arising from the dissipators, which
``push'' population away from the $|M_S=+1\rangle \langle M_S=+1|$ and $|M_S=-1\rangle \langle M_S=-1|$ states into the limit-cycle state 
$|M_S=0\rangle \langle M_S=0|$.
We refer to the system introduced above as the ``ideal spin-1'' (or, more generally, ``ideal spin-$S$'') system.

Throughout this work, we are interested in steady-state phase synchronization  in the presence of a weak external drive, 
quantified by the phase synchronization measure $S_q$~\cite{mok2020},
\begin{eqnarray}
\label{eq_sync}
S_{q} = \left|\sum_{M_S=-S}^{S-1}
\rho_{M_S,M_S+1} \right|.
\end{eqnarray}
Importantly, the coupling strength $\Omega$ of the coherent drive is to be chosen such that the external drive leads to appreciable phase localization but does not break the limit cycle~\cite{koppenhofer2019}.
To visualize phase synchronization, 
it is useful to analyze the Husimi-$Q$ function $Q(\theta,\phi)$~\cite{gilmore1975}, 
\begin{eqnarray}
Q(\theta,\phi)=
\frac{2S+1}{4 \pi}
\langle \theta,\phi| \hat{\rho} | \theta,\phi \rangle,
\end{eqnarray}
where the spin coherent states $|\theta,\phi \rangle$---for each fixed $S$---are constructed by acting 
onto the state $|M_S=S\rangle$,
\begin{eqnarray}
 |\theta,\phi \rangle =
 \exp(-\imath \phi \hat{S}_z)
 \exp(-\imath \theta \hat{S}_y)
 |M_S=S\rangle.
\end{eqnarray}
The Husimi-$Q$ function is normalized such that
$\int_0^{2\pi} \int_0^{ \pi} Q(\theta,\phi) \sin \theta d \theta d \phi=1$.
For integer spin,
the dissipative gain and damping terms both drive population toward the equator ($\theta=\pi/2$) in the $(\theta,\phi)$ phase space. External-drive-induced synchronization, in turn, is accompanied by a non-uniform distribution of the Husimi-$Q$ function in the $\phi$ coordinate, i.e., a maximum of $Q(\theta,\phi)$ for a specific $\phi$ and $\theta \approx \pi/2$. 
Interestingly, steady-state phase localization is absent for the paradigmatic spin-1 system  with equal dissipative gain and damping rates ($\gamma_g=\gamma_d$) for all $\phi_S$. 
    This effect, which is referred to as synchronization blockade~\cite{tan2022,koppenhofer2019,solanki2023}, is due to destructive interference between the
    two immediate off-diagonal density matrix elements $\rho_{-1,0}$ and $\rho_{0,+1}$.

\subsection{Role of auxiliary states}

\label{secII_B}
\begin{figure}[htbp]
{\includegraphics[ scale=.4]{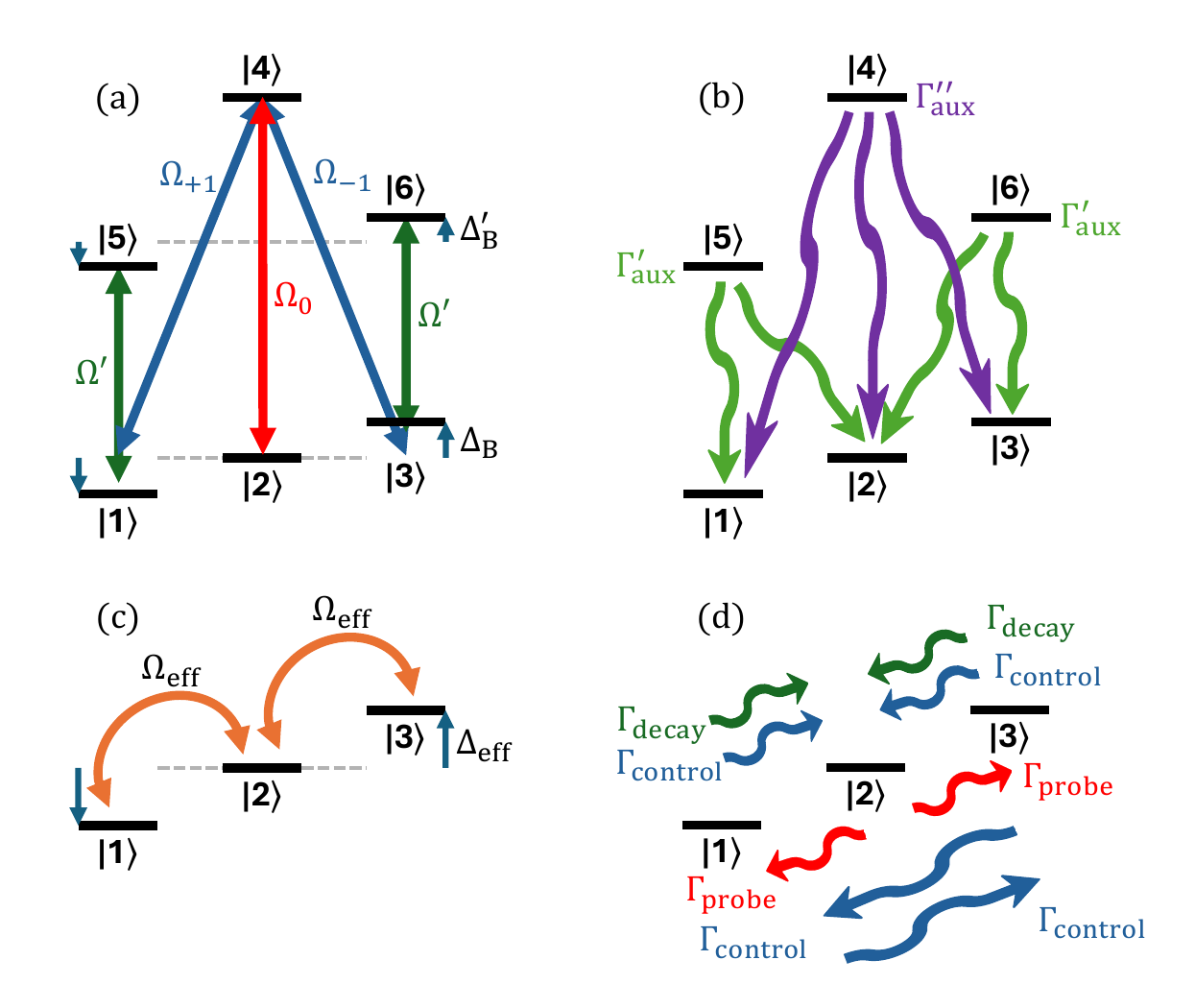}}
\caption{Illustrations of  (a/b) the full $(3+3)$-level system and (c/d) the effective 3-level system.
The black horizontal lines show the energy levels. While the (3+3)-system consists of the ground states $|1\rangle$, $|2\rangle$, and $|3\rangle$
and the auxiliary states $|4\rangle$, $|5\rangle$, and $|6\rangle$, the effective 3-level system only contains the ground states. Gray dashed lines show the energy of the $F=1$ ground state manifold and the $F'=1$ excited state manifold (only the $m_{F'}=\pm1$ states are shown) for vanishing magnetic field.  
Throughout, we assume  $\Delta_\pi''=\Delta_\sigma''=\Delta_\pi'=0$ (these detunings are defined in  Appendix~\ref{sec_appendix0}) and $|\Omega_{+1}|=|\Omega_{-1}|$.
(a) Arrows indicate coherent couplings between  ground and auxiliary states.
Levels $|1\rangle$, $|2\rangle$, and $|3\rangle$ are coupled to $|4\rangle$ with Rabi frequencies $\Omega_{+1}$, $\Omega_{0}$, and $\Omega_{-1}$, respectively. Level $|1\rangle$ is coupled to $|5\rangle$ and level $|3\rangle$ is coupled to $|6\rangle$ with Rabi frequency $\Omega'$. 
The Zeeman shifts for the $F=1$ ground state manifold and the $F'=1$ excited state manifold are $\Delta_B$ and $\Delta_B'$, respectively.
For finite magnetic field, i.e., for non-zero $\Delta_B$ and non-zero $\Delta_B'$, the ``decay'' beams ($\Omega'$) and the ``control'' beams  ($\Omega_{\pm1}$)  are off-resonant while the ``probe'' beam  ($\Omega_{0}$) is on resonance.
The phases of the coherent couplings, which play an important role in our analysis, are not shown.
(b) Wiggly lines show the decay paths of the auxiliary states. The auxiliary state $|4\rangle$ decays into each of $|1\rangle$, $|2\rangle$, and $|3\rangle$ at rate $\Gamma_{\text{aux}}''/3$. The auxiliary state $|5\rangle$ decays into each of $|1\rangle$ and $|2\rangle$ at rate $\Gamma_{\text{aux}}'/2$. The auxiliary state $|6\rangle$ decays into each of $|2\rangle$ and $|3\rangle$ at rate $\Gamma_{\text{aux}}'/2$. 
(c) The two arrows show the coherent effective couplings. Throughout, we work in a parameter regime where the coherent coupling between states $|1\rangle$ and $|3\rangle$ is zero. The effective detuning $\Delta_{\text{eff}}$ comes from the Zeeman shift of the ground state manifold and the light shifts.
(d) The wiggly lines show dissipative effective decay. The control beams ($\Omega_{\pm1}$) mediate effective dissipation from state $|1\rangle$ to states $|2\rangle$ and $|3\rangle$ and from state $|3\rangle$ to states $|1\rangle$ and $|2\rangle$ with rate $\Gamma_{\text{control}}$. The probe beam ($\Omega_{0}$) mediates effective dissipation from state $|2\rangle$ to states $|1\rangle$ and $|3\rangle$ with rate $\Gamma_{\text{probe}}$. The decay beams ($\Omega'$) mediate effective dissipation from states $|1\rangle$ and $|3\rangle$ to state $|2\rangle$ with rate $\Gamma_{\text{decay}}$.  To keep the schematic readable, dissipative effective  self-dephasing processes (``effective self-dissipation'') are not shown (see Appendix~\ref{sec_appendixA}). 
}
\label{steady_states_fig_schematic}
\end{figure}

While the ideal spin-1 system has been studied theoretically in great detail~\cite{roulet2018,roulet2018a,tan2022,koppenhofer2019,solanki2023}, 
 few experimental studies of phase synchronization in isolated driven spin systems exist~\cite{laskar2020,zhang2023,koppenhofer2020}. Moreover, thorough  theoretical benchmark studies, which  account for the auxiliary states that are used in experimental realizations, do not exist.  To fill this gap, 
we consider a $(3+3)$-system that possesses three ground states (this is the Hilbert space of interest) and three auxiliary states, which generate coherent  and  dissipative effective couplings within the ground state manifold.
The chosen states and couplings are applicable to the rubidium $F=1$ hyperfine ground states 
($|1\rangle=|F=1,M_F=+1\rangle$, $|2\rangle=|F=1,M_F=0\rangle$, and $|3\rangle=|F=1,M_F=-1\rangle$);
 the excited $F''=0$ state ($|4\rangle=|F''=0,M_{F''}=0\rangle$); and two excited $F'=1$ states
 ($|5\rangle=|F'=1,M_{F'}=+1\rangle$ and $|6\rangle=|F'=1,M_{F'}=-1\rangle$)~\cite{footnote1}.
 Appendix~\ref{sec_appendix0} reviews the energy level structure of $^{87}$Rb and
 Figs.~\ref{steady_states_fig_schematic}(a) and
 \ref{steady_states_fig_schematic}(b) illustrate the coherent and incoherent couplings, respectively. 
The effective 3-state model, which is obtained by ``integrating out'' the $F''=0$ and $F'=1$ excited states [see Figs.~\ref{steady_states_fig_schematic}(c) and
 \ref{steady_states_fig_schematic}(d) for an illustration of the coherent and incoherent effective couplings, respectively],
is shown below to feature several novel phenomena that are not captured by the paradigmatic spin-1 Hamiltonian, thereby
appreciably advancing our understanding of quantum synchronization in driven spin systems. While our study employs a particular coupling scheme to the auxiliary states, our key findings should apply also to other coupling schemes.

The dynamics of the $(3+3)$-level system in the rotating frame are described by the Lindblad master equation
\begin{align}
\dot{\hat{\rho}} &=
 -i [\hat{H}, \hat{\rho}]
 + \frac{\Gamma_{\text{aux}}''}{3} \sum_{k=1}^{3} 
   \hat{\mathcal{D}}[\hat{L}_{k,4}](\hat{\rho}) \notag \\
&\quad + \frac{\Gamma_{\text{aux}}'}{2} \left(
    \sum_{k=1}^{2} \hat{\mathcal{D}}[\hat{L}_{k,5}](\hat{\rho})
   + \sum_{k=2}^{3} \hat{\mathcal{D}}[\hat{L}_{k,6}](\hat{\rho})
  \right),
\label{steady_states_eq_me_full}
\end{align}
where $\Gamma_{\text{aux}}''$ and $\Gamma_{\text{aux}}'$ encode the finite lifetimes of state
$|4 \rangle$ and states $|5\rangle$ and $|6\rangle$, respectively, 
the fractions of $1/3$ and $1/2$ encode the branching ratios 
for $^{87}$Rb, and $\hat{L}_{k,l}=|k\rangle\langle l|$ are the Lindbladian associated with incoherent decay from state
$|l\rangle$ to state $|k\rangle$.
Fixing the
laser detunings  such that they are zero when the external magnetic field strength is zero (which corresponds to vanishing Zeeman shifts $\Delta_B$ and $\Delta_B'$ of the $F=1$ and $F'=1$ manifolds), the Hamiltonian reads
\begin{widetext}
\begin{eqnarray}
    \label{steady_state_eq_hfull_7by7}
    \underline{H} = \frac{1}{2} 
    \left(
    \begin{array}{ccccccc}
-2\Delta_B
& 0 & 0 &
-|\Omega_{+1}|e^{-i\phi_{+1}} 
& -|\Omega'|e^{-i\phi'} & 0 
\\ 
0 & 0 & 0 &
-|\Omega_{0}|e^{-i \phi_{0}} & 0 & 0 
\\ 
0 & 0 & 2\Delta_B &
-|\Omega_{-1}|e^{-i \phi_{-1}} & 0  &
-|\Omega'|e^{-i\phi'}
\\ 
-|\Omega_{+1}|e^{i\phi_{+1}} &
-|\Omega_{0}|e^{i \phi_{0}} &
-|\Omega_{-1}|e^{i \phi_{-1}} &
0 & 0 & 0 
\\ 
-|\Omega'|e^{i\phi'} & 0 & 0 & 0 &
-2 \Delta_B'  & 0 
\\ 
0 & 0 & -|\Omega'|e^{i\phi'} & 0 & 0  &
2 \Delta_B' 
    \end{array}
    \right) .
\end{eqnarray}
\end{widetext}
The quantities $|\Omega_{+1}|$,  $|\Omega_0|$, $|\Omega_{-1}|$, and $|\Omega'|$ denote magnitudes of Rabi coupling strengths [see Fig.~\ref{steady_states_fig_schematic}(a)] with corresponding phases $\phi_{+1}$, $\phi_0$, $\phi_{-1}$, and $\phi'$. 

To reduce the dynamics to the ground state manifold, we eliminate the excited Hilbert space within the master equation framework, applying the formalism developed in Ref.~\cite{reiter2012} (see Appendix \ref{sec_appendixA} for details). This yields an effective  reduced-Hilbert space Hamiltonian (also referred to as effective low-energy Hamiltonian) that lives in the Hilbert space spanned by the states $|1\rangle$, $|2\rangle$, and $|3\rangle$.  The master equation reduction relies on time-dependent perturbation theory and its validity requires the populations of the excited states to be small.
Enforcing $|\Omega_{+1}|=|\Omega_{-1}|$ (see Appendix~\ref{sec_appendixB} for details), we find that the effective reduced Hilbert-space  Hamiltonian $\hat{H}_{\text{eff}}$
features, just as $\hat{H}_S$, couplings between states $|1\rangle$ and $|2\rangle$ as well as between states
$|2\rangle$ and $|3\rangle$ but not between states $|1\rangle$ and $|3\rangle$,
\begin{eqnarray}
\underline{H}_{\text{eff}} = \nonumber \\
\left(
\begin{array}{ccc}
\Delta_{\text{eff}}
& 
  -i\frac{\Omega_{\text{eff}}}{\sqrt{2}}e^{i(\phi_{\text{eff}}+\pi-\alpha)}
& 
0 
\\ 
  i\frac{\Omega_{\text{eff}}}{\sqrt{2}}e^{-i(\phi_{\text{eff}}+\pi-\alpha)}
&
0
&
-i\frac{\Omega_{\text{eff}}}{\sqrt{2}}e^{i\phi_{\text{eff}}}
\\ 
0 
&
  i\frac{\Omega_{\text{eff}}}{\sqrt{2}}e^{-i\phi_{\text{eff}}}

& 
-\Delta_{\text{eff}}
\end{array}
\right).
\nonumber \\
\label{eq_eff_H_ideal}
\end{eqnarray}
Without finetuning of the laser parameters (see Appendix~\ref{sec_appendixB}), the effective Hamiltonian would possess non-zero coherent coupling between states $|1\rangle$ and $|3\rangle$ due to  second-order processes.  
The effective detuning $\Delta_{\text{eff}}$, effective Rabi coupling strength $\Omega_{\text{eff}}$, and associated phases $\phi_{\text{eff}}$ and $\alpha$ are expressed in terms of the parameters of the $(3+3)$-level system,
\begin{eqnarray}
    \Delta_{\text{eff}}=-\Delta_{B}-\frac{\Delta_{B}|\Omega_{+1}|^2}{(\Gamma_{\text{aux}}'')^2+4\Delta_{B}^2}- \nonumber \\
    \frac{(\Delta_{B}-\Delta_{B}')|\Omega'|^2}{(\Gamma_{\text{aux}}')^2+4(\Delta_{B}-\Delta_{B}')^2} ,
    \label{eq_Delta_eff}
\end{eqnarray}
\begin{equation}
    \Omega_{\text{eff}}=\frac{\sqrt{2}}{8}\frac{|\Delta_{B}||\Omega_{0}||\Omega_{+1}|}{(\Gamma_{\text{aux}}'')^2/4+\Delta_{B}^2}\sqrt{1+\left(\frac{2\Delta_{B}}{\Gamma_{\text{aux}}''}\right)^2},
    \label{eq_Omega_eff}
\end{equation}
\begin{equation}
    \phi_{\text{eff}}=\phi_{-1}-\phi_{0}+\arctan \left(\frac{2\Delta_{B}}{\Gamma_{\text{aux}}''} \right)+\pi-\frac{\pi}{2}\text{sgn}(\Delta_{B}),
    \label{eq_phi_eff}
\end{equation}
and 
\begin{equation}
    \alpha=(\phi_{+1}-\phi_{0})+(\phi_{-1}-\phi_{0}).
\end{equation}
Note that $\Delta_{\text{eff}}$, $\Omega$, $\phi_{\text{eff}}$, and $\alpha$ are all real.
The 
effective 3-level Hamiltonian $\underline{H}_{\text{eff}}$, Eq.~(\ref{eq_eff_H_ideal}), 
is---except for the phase factor $\pi-\alpha$ in the elements $H_{\text{eff},12}$ and $H_{\text{eff},21}$---of 
the same form as the spin-1 Hamiltonian $\underline{H}_S$, Eq.~(\ref{eq_spin1_H}).
Correspondingly, the proposed level scheme realizes the ideal spin-1 system Hamiltonian $\underline{H}_S$
for $\alpha=\pi$.
For $\alpha \ne \pi$, in contrast, our set-up allows---as shown below---for the realization
of synchronization phenomena that are not accessible by the ideal 
spin-1 Hamiltonian. The remainder of this paper works, as enforced when deriving Eq.~(\ref{eq_eff_H_ideal}), with $|\Omega_{+1}|=|\Omega_{-1}|$.

The reduction of the Hilbert space ``converts'' the seven Lindbladian $\hat{L}_{k,l}$ 
into seven effective Lindbladian $\hat{L}_{\text{eff},k,l}$, whose form depends on the coherent couplings and incoherent decay paths of the full Hamiltonian,
\begin{eqnarray}
\label{eq_lindblad_k_4_eff}
\hat{L}_{\text{eff},k,4}
= 
c_{\text{eff},k,1}^{(4)}
|k \rangle \langle 1 |
+ 
c_{\text{eff},k,2}^{(4)}
|k \rangle \langle 2 |
+
c_{\text{eff},k,3}^{(4)}
|k \rangle \langle 3 |
\nonumber \\
\;\;(k=1,2,3),
\end{eqnarray}
\begin{equation}
\label{eq_lindblad_k_5_eff}
\hat{L}_{\text{eff},k,5}
=
c_{\text{eff},k,1}^{(5)}
|k \rangle \langle 1 |
\;\;(k=1,2),
\end{equation}
and 
\begin{equation}
\label{eq_lindblad_k_7_eff}
\hat{L}_{\text{eff},k,6}
=
c_{\text{eff},k,3}^{(6)}
|k \rangle \langle 3 |
\;\;(k=2,3).
\end{equation}
The form of the dissipators $\hat{L}_{\text{eff},k,4}$, namely the fact that they contain a sum of operators, is a bit unusual. The sum emerges as part of  the master equation reduction since state $|4\rangle$ is coupled coherently to more than one state~\cite{reiter2012}. 
While the Lindbladian $\hat{L}_{k,l}$ describes dissipative processes that start at the excited state $|l\rangle$ and end at the ground state
$|k\rangle$ (in our case, incoherent decay due to the finite excited state lifetime), the effective Lindbladian 
$\hat{L}_{\text{eff},k,l}$ describes effective processes that start, in general, at any of the ground states $|j\rangle$, 
proceed via the
excited state $|l\rangle$, and end at the ground state $|k\rangle$.
The associated prefactors $c_{\text{eff},k,j}^{(l)}$ are given by \begin{equation}
    \label{steady_states_eq_lindblad_eff_gamma}
    c_{\text{eff},k,j}^{(l)}
    = \frac{\sqrt{\Gamma_{k,l}} H_{l,j}}{ H_{l,l} - H_{j,j}-i \Gamma_{l}/2 },
\end{equation}
where 
$\Gamma_{l}$
is the total
decay rate out of $|l\rangle$ (in our case, $\Gamma_4=\Gamma_{\text{aux}}''$,
$\Gamma_5=\Gamma_6=\Gamma_{\text{aux}}'$) and
$\Gamma_{k,l}$ is the decay rate from $|l\rangle$ to $|k\rangle$, which satisfies $\sum_{k=1}^{3}\Gamma_{k,l}=\Gamma_{l}$  
(in our case, $\Gamma_{k,4}=\Gamma_{\text{aux}}''/3$, $\Gamma_{1,5}=\Gamma_{2,5}=\Gamma_{2,6}=\Gamma_{3,6}=\Gamma_{\text{aux}}'/2$, and $\Gamma_{3,5}=\Gamma_{1,6}=0$).
The quantity $|c_{\text{eff},k,j}^{(l)}|^{2}$ corresponds to the effective rate 
for going from state $|j\rangle$ to state $|k\rangle$ via the auxiliary state
$|l\rangle$.
With these definitions, the effective master equation assumes the ``standard form,'' namely~\cite{reiter2012}
\begin{eqnarray}
\label{steady_states_eq_me_full}
\dot{\hat{\rho}}_{\text{eff}} 
=
 -i [\hat{H}_{\text{eff}}, \hat{\rho}_{\text{eff}}]  + \sum_{k=1}^{3} \hat{\mathcal{D}}[\hat{L}_{\text{eff},k,4}](\hat{\rho}) 
 \nonumber 
 \\
+ \sum_{k=1}^{2} \hat{\mathcal{D}}[\hat{L}_{\text{eff},k,5}](\hat{\rho})   + \sum_{k=2}^{3} \hat{\mathcal{D}}[\hat{L}_{\text{eff},k,6}](\hat{\rho}).
\end{eqnarray}
It should be noted that our convention is such that $\hat{L}_{k,l}$ is dimensionless while $\hat{L}_{\text{eff},k.l}$ carries units. The motivation for this convention 
is that the prefactors $c_{\text{eff},k,l}^{(4)}$ in Eq.~(\ref{eq_lindblad_k_4_eff}) cannot be pulled out, i.e., 
 in general, a common multiplicative factor that can be pulled out of the dissipator does not exist. 
As shown in the next section, this  characteristic distinguishes the effective
3-level master equation from the master equation for the ideal spin-1 system.
 In particular, since the prefactor of each of the three operators on 
 the right hand side of Eq.~(\ref{eq_lindblad_k_4_eff})  
 is, in general, a complex number,  
  the non-linearity of the dissipator leads to interferences 
that may play a critical role in the time evolution. 
 Motivated by the interpretation of the ``standard'' master equation, we refer to couplings that arise from the commutator and dissipators on the right hand side of Eq.~(\ref{steady_states_eq_me_full}) as  coherent effective couplings and  incoherent effective couplings, respectively. Since the incoherent effective couplings may depend on the phases of the coherent couplings of the full $(3+3)$-level system, it can be expected that the   dissipative effective couplings lead to physics that is not captured by the dissipators of the ideal spin-1 system. The next section investigates this physics.

\section{Results}
\label{secIII}
The previous section reduced the master equation for the (3+3)-level system to an effective
master equation for a 3-level system whose system Hamiltonian is,  for specific parameter combinations,  equivalent to that
of the ideal spin-1 Hamiltonian and whose Lindbladian have a more complicated structure than those of the 
ideal spin-1 system. Throughout this result section, 
we work in parameter regimes where the effective 3-level  master equation provides a faithful description of the
dynamics of the full (3+3)-level system, i.e, we focus
on parameter combinations for which
$\langle j |\hat{\rho}_{\text{eff}} | k \rangle \approx \langle j |\hat{\rho} | k \rangle$ for $j,k=1,2,3$.
As a consequence,  the synchronization $S_q$, calculated using the 12- and 23-matrix elements of $\hat{\rho}$, is very similar to that calculated using 
the 12- and 23-matrix elements of $\hat{\rho}_{\text{eff}}$ [see Eq.~(\ref{eq_sync})].  Throughout, analytical first-order
perturbation theory expressions, derived from the effective 3-level master equation (see Appendixes~\ref{sec_appendixBpert} and \ref{sec_appendixD} for details), are used to interpret our results.  
Our analysis is divided into two parts:
Section~\ref{secIIIA} considers systems with $\Delta_B \ne 0$ while Sec.~\ref{secIIIB} 
considers systems with $\Delta_B=0$ (recall that
$\Delta_B=0$ implies $ \Delta_{\text{eff}} = \Omega_{\text{eff}}=0$).  Sections~\ref{secIIIA} and \ref{secIIIB} both use
decay rates $\Gamma_{\text{aux}}'$ and $\Gamma_{\text{aux}}''$ applicable to $^{87}$Rb  (see Appendix~\ref{sec_appendix0}) and $|\Omega_{+1}|=|\Omega_{-1}|$.

\subsection{Finite Zeeman splitting: $\Delta_{B} \ne 0$}
\label{secIIIA}
For finite $\Delta_B$, the steady-state solution in the absence of an external drive (i.e., for $\Omega_0=0$)
is equal to $|2\rangle \langle 2|$ for both the full and effective master equations (see Appendix~\ref{sec_appendixC} for details). Since the steady-state solution is independent of the initial conditions, $|2\rangle \langle 2|$ is a limit cycle state. Steady-state phase synchronization of the effective spin-1 system to the external drive is realized when the majority of the population  in the presence of the external drive is in 
the limit cycle state, with the phase localizing at a specific $\phi$-value.

\begin{figure}[t]
{\includegraphics[scale=.33]{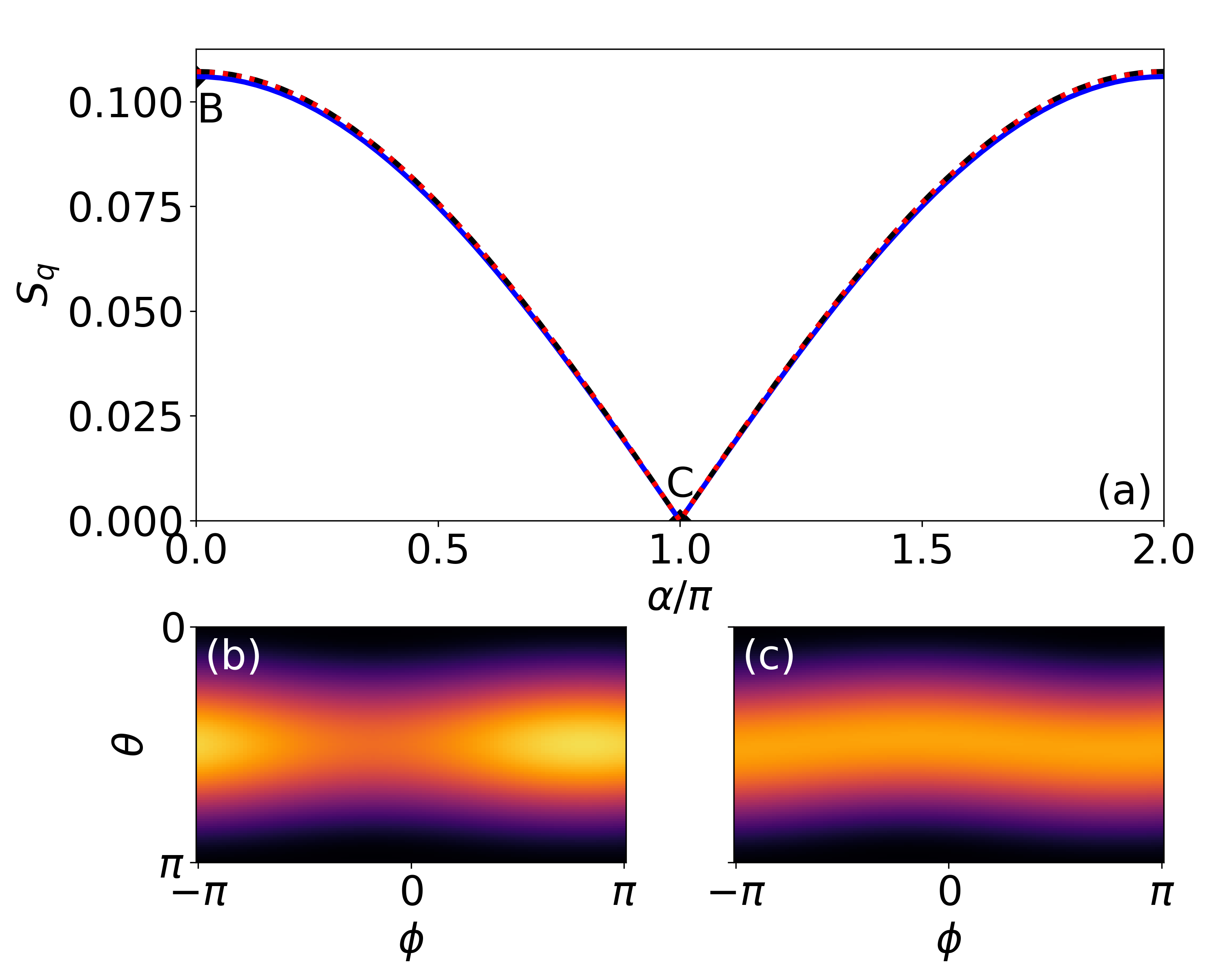}}
\caption{(a) Steady-state synchronization $S_q$ as a function of the phase angle $\alpha$ for $\Delta_B=2\pi \times 0.4$~MHz, $\phi_{\pm1}=\phi_0=\phi'=0$, $|\Omega_{\pm1}|=2\pi \times 9.5$~MHz, $|\Omega_{0}|=2\pi \times 1.0$~MHz, and  $|\Omega'|=2\pi \times 3.0$~MHz (the decay rates $\Gamma_{\text{aux}}'$ and $\Gamma_{\text{aux}}''$ are those for $^{87}$Rb; see Appendix~\ref{sec_appendix0}). The  black dashed, blue solid, and red dotted lines are for the full $(3+3)$-level system, the effective 3-level model, and the perturbative treatment of the effective 3-level system, respectively. The three curves nearly coincide for all $\alpha$. Husimi-$Q$ distributions for points B ($\alpha=0$) and C ($\alpha=\pi$) as functions of $\phi$ (horizontal axis) and $\theta$ (vertical axis) are shown in (b) and (c), respectively; the color scheme is the same as that used in Figs.~\ref{rewrite_fig_beta} and \ref{rewrite_fig_HusimiQ}. The Husimi-$Q$ function for $\alpha=0$ shows phase localization while that for $\alpha=\pi$ shows a small deformation of the limit cycle state but no phase localization.} 
\label{rewrite_fig_alpha}
\end{figure}

Figures~\ref{rewrite_fig_alpha}(b) and \ref{rewrite_fig_alpha}(c) show steady-state Husimi-$Q$ functions for two values of $\alpha$ ($\alpha=\pi$ and $\alpha=0$) 
for  $\Delta_B=2 \pi \times 0.4$~MHz, $|\Omega_{\pm1}|=2 \pi \times 9.5$~MHz,
 $|\Omega_{0}|=2 \pi \times 1$~MHz, and $|\Omega'|=2 \pi \times 3$~MHz.
For $\alpha =\pi$ (i.e., when $\hat{H}_{\text{eff}}$ is equal to $\hat{H}_S$),
the Husimi-$Q$ function of the effective spin-1 system corresponds to a slightly deformed limit cycle that is characterized, just as the ideal spin-1 system with the simpler dissipators, by vanishing synchronization. Specifically, the integral  $\int_{-1}^{1}Q(\theta,\phi)d\cos \theta$  is the same for all $\phi$. This shows that the more complicated dissipator structure of the effective 3-level system
does not remove the synchronization blockade observed in the literature  for the ideal spin-1 system~\cite{tan2022}. 
For $\alpha=0$, in contrast, the steady-state Husimi-$Q$ function has a global maximum at $(\theta_{\text{max}},\phi_{\text{max}})\approx (\pi/2,0.815\pi)$, which signals that the phase synchronization of the driven effective spin-1 system is finite. Note that the value of $\phi_{\text{max}}$ can be controlled by changing the phases of the Rabi couplings (see also below).  We conclude from  Figs.~\ref{rewrite_fig_alpha}(b) and \ref{rewrite_fig_alpha}(c) that the synchronization blockade can be removed by varying the phase angle $\alpha$. This important result agrees with Ref.~\cite{laskar2020}, which pursued  an {\em{ad hoc}} master equation reduction as opposed to a systematic elimination of the excited states, as done in the present work. 

Black dashed, blue solid, and red dotted lines in Fig.~\ref{rewrite_fig_alpha}(a)
show $S_q$ for the full (3+3)-level system, the effective 3-level system, and 
the perturbative treatment of the effective 3-level system.
The excellent  agreement between these three descriptions validates the use of the effective 3-level master equation and indicates that the perturbative framework provides a robust tool for gaining physical insights into the system behavior.
It can be seen  that $S_q$ takes on a minimum for $\alpha= n \pi$, where $n=\pm 1,\pm 3,\dots$, and
a maximum for $\alpha=n \pi $, where $n=0, \pm2,\pm4,\dots$. Together, Fig.~\ref{rewrite_fig_alpha} shows that the phase angle $\alpha$, which is controlled by the phases of the coherent couplings that realize the external drive, allows one to tune the synchronization of the effective spin-1 system, which is characterized by seven effective dissipators.
Due to our choice of the detunings and due to working with $|\Omega_{+1}|=|\Omega_{-1}|$
, the magnitudes of the rates 
that describe decay from state $|1\rangle$ to state $|2\rangle$ and from state $|3\rangle$ to state $|2\rangle$ are equal to each other. In the ideal spin-1 system, this would correspond to $\gamma_g=\gamma_d$ [see Eq.~(\ref{eq_me_standard})] and would imply vanishing synchronization due to the synchronization blockade. For the effective model, the synchronization blockade can be avoided by tuning $\alpha$, i.e., by taking advantage of the ``unusual'' functional form of $\hat{L}_{\text{eff},k,4}$ (see the discussion in Sec.~\ref{secII_B}).

To gain insights into how the $\alpha$-dependence emerges and, more generally, to better understand which terms in the effective master equation ``generate'' the finite synchronization, we
analyze the system behavior using first-order perturbation theory.
To simplify the notation, we  define
$|c_{k,1}^{(4)}|^2=|c_{k,3}^{(4)}|^2=\Gamma_{\text{control}}$ ($k=1,2,3$); $|c_{k,2}^{(4)}|^2=\Gamma_{\text{probe}}$ ($k=1,2,3$); $|c_{k,1}^{(5)}|^2=\Gamma_{\text{decay}}$ ($k=1,2$); and $|c_{k,3}^{(6)}|^2=\Gamma_{\text{decay}}$ ($k=2,3$). 
Performing first-order perturbation theory around the limit cycle state $|2\rangle \langle 2|$ (see Appendix~\ref{sec_appendixD} for details), the steady-state synchronization of the effective master equation reads 
\begin{widetext}
\begin{eqnarray}
\label{eq_sync_pt_finite}
S_q=\left| \cos \left(
\frac{\alpha}{2} \right) \right| \frac{\left|2\left|H_{\text{eff,2,3}}\right|\left(\left|\Delta_{\text{eff}}\right|-i\Gamma_{\text{decay}}\right)+3i\Gamma_{\text{control}}^{1/2}\Gamma_{\text{probe}}^{1/2}\left(\left|\Delta_{\text{eff}}\right|-i\Gamma_{\text{decay}}\right)-6i\left|H_{\text{eff},2,3}\right|\Gamma_{\text{control}}\right|}{\Delta_{\text{eff}}^2+3\Gamma_{\text{decay}}\Gamma_{\text{control}}+\Gamma_{\text{decay}}^2}.
\end{eqnarray}
\end{widetext}
The three terms in the  
numerator in Eq.~(\ref{eq_sync_pt_finite}) 
represent contributions to the synchronization due to coherent effective couplings, due to incoherent effective  couplings, and due to an interplay of coherent and incoherent couplings. Since the terms 
in the numerator
in Eq.~(\ref{eq_sync_pt_finite}) do not depend on any phases, the only phase dependence of the perturbative $S_q$ expression is due to the $|\cos(\alpha/2)|$ factor. 
Equation~(\ref{eq_sync_pt_finite})  explains the  $\alpha$-dependence of $S_q$ observed in Fig.~\ref{rewrite_fig_alpha}. The dependence of $S_q$ on the relative phase $\alpha$ is reminiscent of  standard  
STIRAP set-ups~\cite{vitanov2017}, where the dynamics is governed by the relative laser phase, as well as spin-1 Bose-Einstein condensates~\cite{stamperkurn2013,kawaguchi2012}, where the mean-field dynamics is governed by the relative  phase of the three spin components.

\begin{figure}[h]
{\includegraphics[scale=.4]{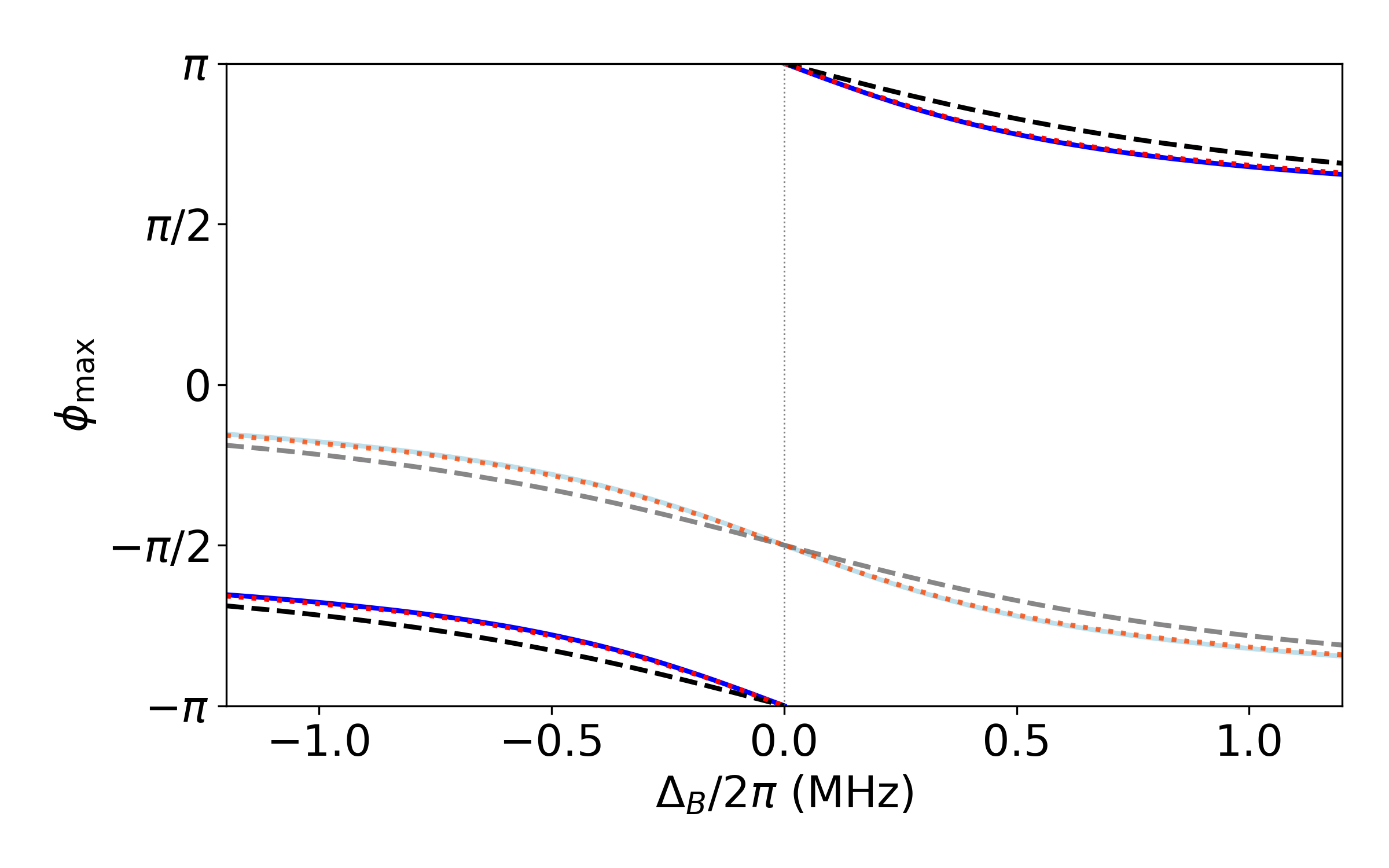}}
\caption{Phase angle $\phi_{\text{max}}$ at which the Husimi-$Q$ function takes its maximum as a function of $\Delta_B$ for $\phi_0=0$, $\phi'=0$, $|\Omega_{\pm1}|=2\pi \times 9.5$~MHz, $|\Omega_{0}|=2\pi \times 1.0$~MHz, and  $|\Omega'|=2\pi \times 3.0$~MHz (the decay rates $\Gamma_{\text{aux}}'$ and $\Gamma_{\text{aux}}''$ are those for $^{87}$Rb; see Appendix~\ref{sec_appendix0}).
Two different parameter combinations of the control laser phases are considered:
The dark lines are for $\phi_{\pm 1} = 0$ while the faded lines are for $\phi_{\pm 1} = \pm \pi/2$.
The dark/faded black dashed, dark/faded blue solid, and dark/faded red dotted lines show results for the full $(3+3)$-level system, the effective 3-level system, and the perturbative treatment of the effective 3-level system, respectively.}
\label{rewrite_fig_phi_max}
\end{figure}

While the phase angle $\alpha$ governs $S_q$ at first-order perturbation theory, an analogous analysis (we do not show the equation) shows that the phase angle $\phi_{\text{max}}$ at which the Husimi-$Q$ function takes its maximum does not only depend on the overall phase $\alpha$ but also on other relative laser phases.
To illustrate this, dark and faded lines in Fig.~\ref{rewrite_fig_phi_max} show $\phi_{\text{max}}$ for $\phi_0=0$ as a function of the Zeeman splitting $\Delta_B$ for $\phi_{\pm 1} = 0$ and $\phi_{\pm 1} = \pm \pi/2$, respectively. 
For the parameters considered in Fig.~\ref{rewrite_fig_phi_max}, the angle $\theta$ at which the Husimi-$Q$ function takes its maximum is approximately equal to $\pi/2$. Even though both phase combinations considered in Fig.~\ref{rewrite_fig_phi_max} correspond to the same value of $\alpha$ (namely, $\alpha=0$), the value of $\phi_{\text{max}}$ for fixed $\Delta_B$ is different, indicating that $\phi_{\text{max}}$ is not fully governed by the phase angle $\alpha$.

In what follows, we focus on the phase angle $\alpha=0$, which yields the largest synchronization (see Fig.~\ref{rewrite_fig_alpha}). 
Our aim is to contrast the behaviors of the effective 3-level system with $\alpha=0$ and the behaviors of an ``expanded ideal spin-1 model'' that has the same dissipators as the ideal spin-1 model (dissipative rates $\gamma_g$ and $\gamma_d$) but whose Hamiltonian $\underline{H}_{S,\text{exp}}$ contains, just as the effective 3-level Hamiltonian with $\alpha=0$, an additional phase factor on one of the off-diagonal couplings,
\begin{equation}
\label{eq_spin1_H_matrix}
    \underline{H}_{S,\text{exp}}
    =\left(
    \begin{array}{ccc}
    \Delta
&
-i \frac{\Omega e^{i(\phi_{S}+\pi)}}{\sqrt{2}}
&
0
\\
i \frac{\Omega e^{-i(\phi_{S}+\pi)}}{\sqrt{2}}
&
0
&
-i \frac{\Omega e^{i\phi_{S}}}{\sqrt{2}}
\\
0
&
i \frac{\Omega e^{-i\phi_{S}}}{\sqrt{2}}
&
-\Delta
    \end{array}
    \right).
\end{equation}
To unravel the dissipator structure of the effective 3-level system, we introduce an {\em{ad hoc}} scaling factor $\beta$ ($0 \le \beta \le 1$) into the effective Lindbladians following two different approaches. In both approaches, $\beta=1$ corresponds to the effective 3-level system while $\beta<1$ modifies the system in an {\em{ad hoc}} manner. The aim is to track the synchronization characteristics as $\beta$ is reduced from $1$ (this describes the effective 3-level system) to $0$ (this describes a simpler system that can be connected or contrasted with the expanded ideal spin-1 system).

Approach (1):
The coefficients $c_{\text{eff},k,1}^{(5)}$ ($k=1$), $c_{\text{eff},k,3}^{(6)}$ ($k=3$), and 
$c_{\text{eff},k,j}^{(4)}$ ($k,j=1,2,3$) are multiplied by $\beta$. In this approach, the scaling factor $\beta$ is introduced such that the dissipative terms of the effective 3-level system reduce to those of the ideal spin-1 model for $\beta=0$, i.e., we can identify 
$\gamma_g=\gamma_d=\Gamma_{\text{decay}}$ for $\beta=0$. Recall that $\Gamma_{\text{decay}}$ characterizes effective dissipative coupling from state $|1\rangle$ to $|2\rangle$ as well as from state $|3\rangle$ to state $|2\rangle$ (see Fig.~\ref{steady_states_fig_schematic}). 
The synchronization $S_{q,a1}$ for approach (1), at the level of first-order perturbation theory,
 reads
\begin{widetext}
\begin{eqnarray}
\label{eq_sync_pt_finite_beta1}
S_{q,a1}= \left| \cos \left( \frac{\alpha}{2} \right) \right| \times \nonumber \\
\frac{\left|2\left|H_{\text{eff,2,3}}\right|\left(\left|\Delta_{\text{eff}}\right|-i\frac{1+\beta^2}{2}\Gamma_{\text{decay}}\right)+3i\beta^2\Gamma_{\text{control}}^{1/2}\Gamma_{\text{probe}}^{1/2}\left(\left|\Delta_{\text{eff}}\right|-i\frac{1+\beta^2}{2}\Gamma_{\text{decay}}\right)-6i\beta^2\left|H_{\text{eff},2,3}\right|\Gamma_{\text{control}}\right|}{\Delta_{\text{eff}}^2+3\beta^2\frac{1+\beta^2}{2}\Gamma_{\text{decay}}\Gamma_{\text{control}}+\left( \frac{1+\beta^2}{2}\right)^2 \Gamma_{\text{decay}}^2}.
\end{eqnarray}
\end{widetext}
Equation~(\ref{eq_sync_pt_finite_beta1}) reduces, as it should, to Eq.~(\ref{eq_sync_pt_finite}) for $\beta=1$ and  to
\begin{eqnarray}
\label{eq_sync_pt_finite_beta1_equals_0}
S_{q,a1}=\left| \cos \left( \frac{\alpha}{2} \right) \right| 
\frac{\left|2 H_{\text{eff,2,3}} \left(\left|\Delta_{\text{eff}}\right|-i\frac{\Gamma_{\text{decay}}}{2}\right)\right|}{\Delta_{\text{eff}}^2+\frac{1}{4}\Gamma_{\text{decay}}^2}
\end{eqnarray}
for $\beta=0$. For $|\alpha| \ne \pi, 2 \pi,\dots$,  Eq.~(\ref{eq_sync_pt_finite_beta1_equals_0}) yields, in general, a finite synchronization; this is due to the $\pi$-phase factor 
in the effective 3-level system. This $\pi$-phase factor is present 
in the expanded ideal spin-1 system but not in the ``usual'' ideal spin-1 model. Specifically, the first-order perturbation theory expression yields $S_q=0$ for the ideal spin-1 system (Hamiltonian $\underline{H}_{S}$) but, in general, $S_q \ne 0$ for the expanded spin-1 system (Hamiltonian $\underline{H}_{S,\text{exp}}$). 

Approach (2): The coefficients  
$c_{\text{eff},k,1}^{(5)}$ ($k=1$), $c_{\text{eff},k,3}^{(6)}$ ($k=3$), 
$c_{\text{eff},k,2}^{(4)}$ ($k=1,2,3$), 
$c_{\text{eff},k,1}^{(4)}$ ($k=1,3$), and 
$c_{\text{eff},k,3}^{(4)}$ ($k=1,3$) are multiplied by $\beta$.
In this case, the scaling factor $\beta$ is introduced such that for $\beta=0$ the dissipative effective processes that push population into state $|2\rangle \langle 2|$ (governed by the decay and control beams) survive while those that push population out of state $|2\rangle \langle 2|$ (governed by the probe beam) are turned off.
The synchronization $S_{q,a2}$ for approach (2), at the level of first-order perturbation theory, reads 
\begin{widetext}
\begin{eqnarray}
\label{eq_sync_pt_finite_beta2}
S_{q,a2}= \left| \cos \left( \frac{\alpha}{2} \right) \right| \times \nonumber \\
\frac{\left|2\left|H_{\text{eff,2,3}}\right|\left(\left|\Delta_{\text{eff}}\right|-i\frac{1+\beta^2}{2}\Gamma_{\text{decay}}\right)+i(\beta+2\beta^2)\Gamma_{\text{control}}^{1/2}\Gamma_{\text{probe}}^{1/2}\left(\left|\Delta_{\text{eff}}\right|-i\frac{1+\beta^2}{2}\Gamma_{\text{decay}}\right)-2i(1+2\beta^2)\left|H_{\text{eff},2,3}\right|\Gamma_{\text{control}}\right|}{\Delta_{\text{eff}}^2+(1+2\beta^2)\frac{1+\beta^2}{2}\Gamma_{\text{decay}}\Gamma_{\text{control}}+\left(\frac{1+\beta^2}{2}\right)^2\Gamma_{\text{decay}}^2}. \nonumber \\
\end{eqnarray}
\end{widetext}
Equation~(\ref{eq_sync_pt_finite_beta2}) reduces, as it should, to Eq.~(\ref{eq_sync_pt_finite}) for $\beta=1$ and to
\begin{eqnarray}
\label{eq_sync_pt_finite_beta2_equals_0}
S_{q,a2}= \left| \cos \left( \frac{\alpha}{2} \right) \right| \times \nonumber \\
\frac{\left|2\left|H_{\text{eff,2,3}}\right|\left(\left|\Delta_{\text{eff}}\right|-i\frac{\Gamma_{\text{decay}}}{2}\right)-2i\left|H_{\text{eff,2,3}}\right|\Gamma_{\text{control}}\right|}{\Delta_{\text{eff}}^2+\frac{1}{2}\Gamma_{\text{control}}\Gamma_{\text{decay}}+\frac{1}{4}\Gamma_{\text{decay}}^2}
\end{eqnarray}
for $\beta=0$.
Since the dissipative terms for $\beta=0$ do not reduce to those of the expanded ideal spin-1 model,
i.e., 
the dissipative terms for $\beta=0$  cannot be reduced to 
$\gamma_g {\hat{\cal{D}}}[\hat{L}_g](\hat{\rho})+\gamma_d {\hat{\cal{D}}}[\hat{L}_d](\hat{\rho})$,
$S_{q,a2}$ for the effective model with $\beta=0$ is not the same as $S_q$ for the expanded ideal spin-1 system. 
Specifically, there exists an additional term in $S_{q,a2}$ for $\beta=0$ that depends on $\Gamma_{\text{control}}$.
This term exists since the effective Lindbladian
$\hat{L}_{\text{eff},2,4}=c_{\text{eff},2,1}^{(4)}|2\rangle \langle 1|+  c_{\text{eff},2,3}^{(4)}|2\rangle \langle 3|$  (the term proportional to $c_{\text{eff},2,2}^{(4)}$ goes to zero for $\beta=0$)
 depends on the relative phase between the coefficients
$c_{\text{eff},2,1}^{(4)}$ and $c_{\text{eff},2,3}^{(4)}$.

\begin{figure}[h]
{\includegraphics[scale=.325]{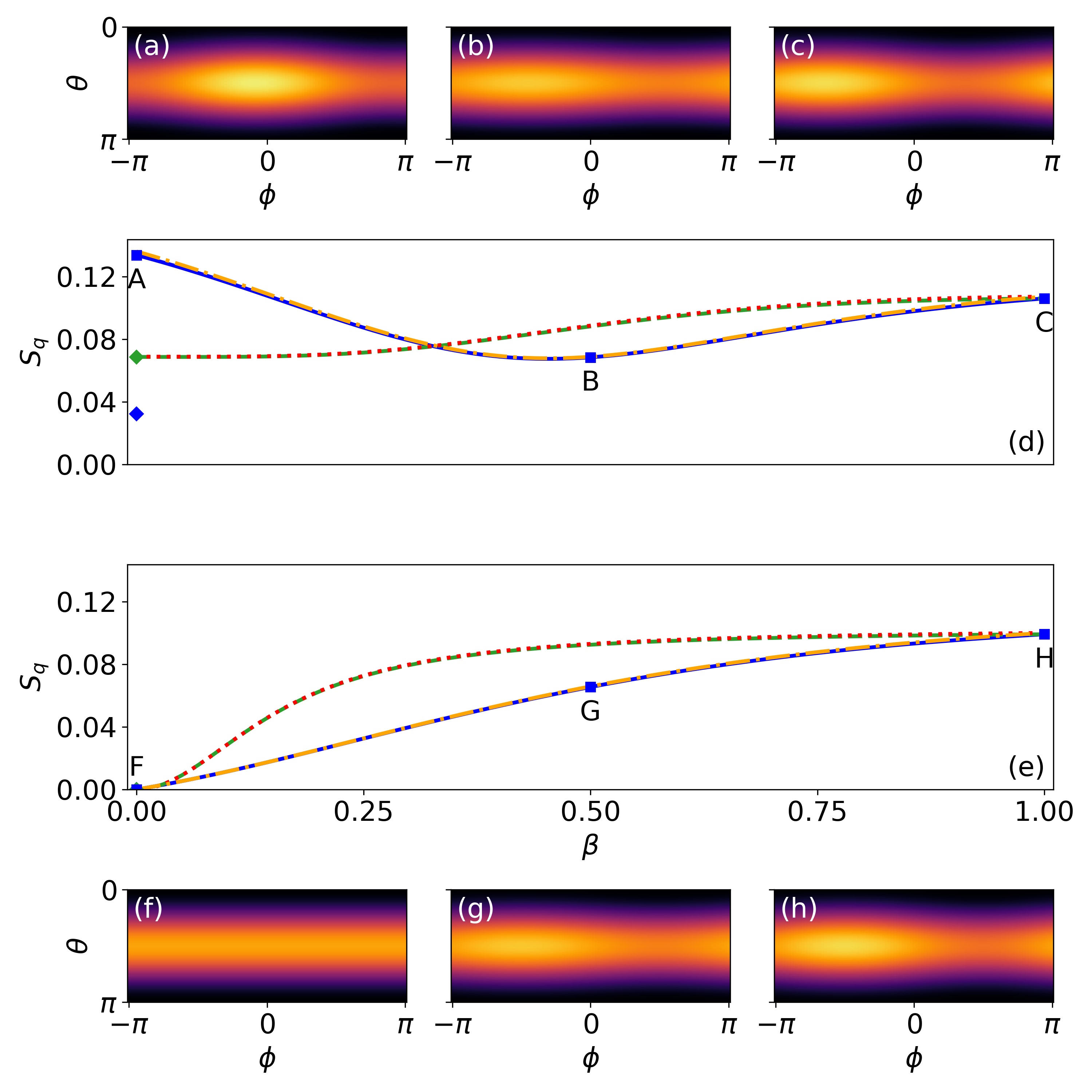}}
\caption{
The green dashed and blue solid lines in panels (d) and (e) show the synchronization $S_q$, calculated by solving the master equation numerically for the effective 3-level system with scaled dissipative rates (see text for details), as a function of $\beta$ for approach (1) and approach (2), respectively. 
Panel
(d) is for $\Delta_B=2\pi \times 0.4$~MHz and panel (e) for $\Delta_B=0$;
the other parameters are the same as in Fig.~\ref{rewrite_fig_alpha}, except for $\phi_{-1}$, which is chosen such that $\phi_{\text{eff}}=0$ holds. 
For comparison, the red dotted and orange dash-dotted lines show the corresponding perturbative expressions $S_{q,a1}$ [Eq.~(\ref{eq_sync_pt_finite_beta1})]
and $S_{q,a2}$ [Eq.~(\ref{eq_sync_pt_finite_beta2})], respectively.
 Husimi-$Q$ distributions for point A ($\beta=0$), point B ($\beta=1/2$), and point C ($\beta=1$) are included above panel (d) for $\Delta_B=2 \pi \times 0.4$~MHz while 
Husimi-$Q$ distributions for point F ($\beta=0$), point G ($\beta=1/2$), and point H ($\beta=1$) are included below panel (e) for $\Delta_B=0$; the color scheme is the same as that used in Figs.~\ref{rewrite_fig_alpha} and \ref{rewrite_fig_HusimiQ}.  
The green and blue diamonds at $\beta=0$ show $S_q$ for the expanded spin-1 model using $\gamma_g=\gamma_d=\Gamma_{\text{decay}}$ and
$\gamma_g=\gamma_d=\Gamma_{\text{decay}}+\Gamma_{\text{control}}$, respectively.
The parameters for the expanded spin-1 system follow from Eqs.~(\ref{eq_Delta_eff}),(\ref{eq_Omega_eff}),(\ref{eq_phi_eff}), and (\ref{steady_states_eq_lindblad_eff_gamma}).
}
\label{rewrite_fig_beta}
\end{figure}

The green dashed and blue solid lines in Fig.~\ref{rewrite_fig_beta}(d) show  $S_q$ as a
function of the {\em{ad hoc}}  scaling parameter $\beta$ for approaches (1) and (2), respectively, for finite $\Delta_B$. The results are obtained by solving the effective 3-level master equation numerically for  $\alpha=0$ (the other parameters are the same as those used in Fig.~\ref{rewrite_fig_alpha}). 
For comparison, the red dotted and orange dash-dotted lines show the corresponding perturbative expressions $S_{q,a1}$ and $S_{q,a2}$, respectively, as a function of $\beta$. The excellent agreement between the perturbative results and the results for the effective 3-level master equation indicates that the effective 3-level results can be interpreted within the framework of first-order perturbation theory. Moreover, since the results for the effective 3-level system agree excellently with the results for the full $(3+3)$-level system for $\beta=1$, the perturbative treatment of the effective 3-level system provides a reliable tool for understanding the behavior of the full $(3+3)$-level system.

For approach (1), $S_q$ decreases monotonically with decreasing $\beta$ and reaches, for $\beta=0$, the same value as the synchronization $S_q$ of the expanded spin-1 model with the same Hamiltonian and dissipative rates $\gamma_g=\gamma_d=\Gamma_{\text{decay}}$.  
Figure~\ref{rewrite_fig_beta}(d) highlights an important result, namely that $S_q$ can, for an identical $3 \times 3$ Hamiltonian, be larger for the more complicated dissipators of the effective model [Eqs.~(\ref{eq_lindblad_k_4_eff})-(\ref{steady_states_eq_me_full})] than for the ``conventional'' dissipators of the expanded ideal spin-1 system [Eq.~(\ref{eq_me_standard})]. The enhancement of the synchronization is due to the fact that the more complicated dissipators contain terms that dissipatively push population out of the limit cycle state, in addition to containing terms that dissipatively push population into the limit cycle state. We will return to this point in the next section, which discusses the  $\Delta_B=0$ case.

For approach (2), $S_q$ decreases monotonically with decreasing $\beta$ for $1 \ge \beta \gtrsim 0.5$
and then increases again for $0.5 \gtrsim \beta \ge 0$. The non-monotonic dependence of $S_{q,a2}$ on $\beta$ is due to the interplay of the three terms inside the second absolute value signs in Eq.~(\ref{eq_sync_pt_finite_beta2}). Using the parameters from Table~\ref{tab:deltaB_parameters}, one can show, e.g., that the minimum of $S_{q,a2}$ at $\beta\approx 1/2$ is due to the near cancellation of the imaginary parts of the first, second, and third terms.  The synchronization for $\beta=0$ is enhanced due to the phase
dependence of the dissipative terms [second term in the numerator of Eq.~(\ref{eq_sync_pt_finite_beta2_equals_0})]. 
Notably, for $\beta=0$, the synchronization $S_q$ of the expanded spin-1 model with the same Hamiltonian and dissipative rates $\gamma_g=\gamma_d=\Gamma_{\text{decay}}+\Gamma_{\text{control}}$ [blue diamond in Fig.~\ref{rewrite_fig_beta}(d)] yields a much lower synchronization than the {\em{ad hoc}} model for all $\beta < 1$ as well as for the ``true'' $\beta=1$ system. This shows that the dissipative terms that push population into the limit cycle state $|2\rangle \langle 2|$ play an important role in synchronizing the system. 
Figures~\ref{rewrite_fig_beta}(a), \ref{rewrite_fig_beta}(b), and \ref{rewrite_fig_beta}(c) show Husimi-$Q$ functions for points A, B, and C, respectively, demonstrating phase localization.  The change of $\phi_{\text{max}}$ with $\beta$ is captured by the first-order perturbation theory expressions (not shown).

\subsection{Zero Zeeman splitting: $\Delta_{B} = 0$}
\label{secIIIB}

\begin{figure}[h]
\vspace*{0.2in}
{\includegraphics[scale=0.75]{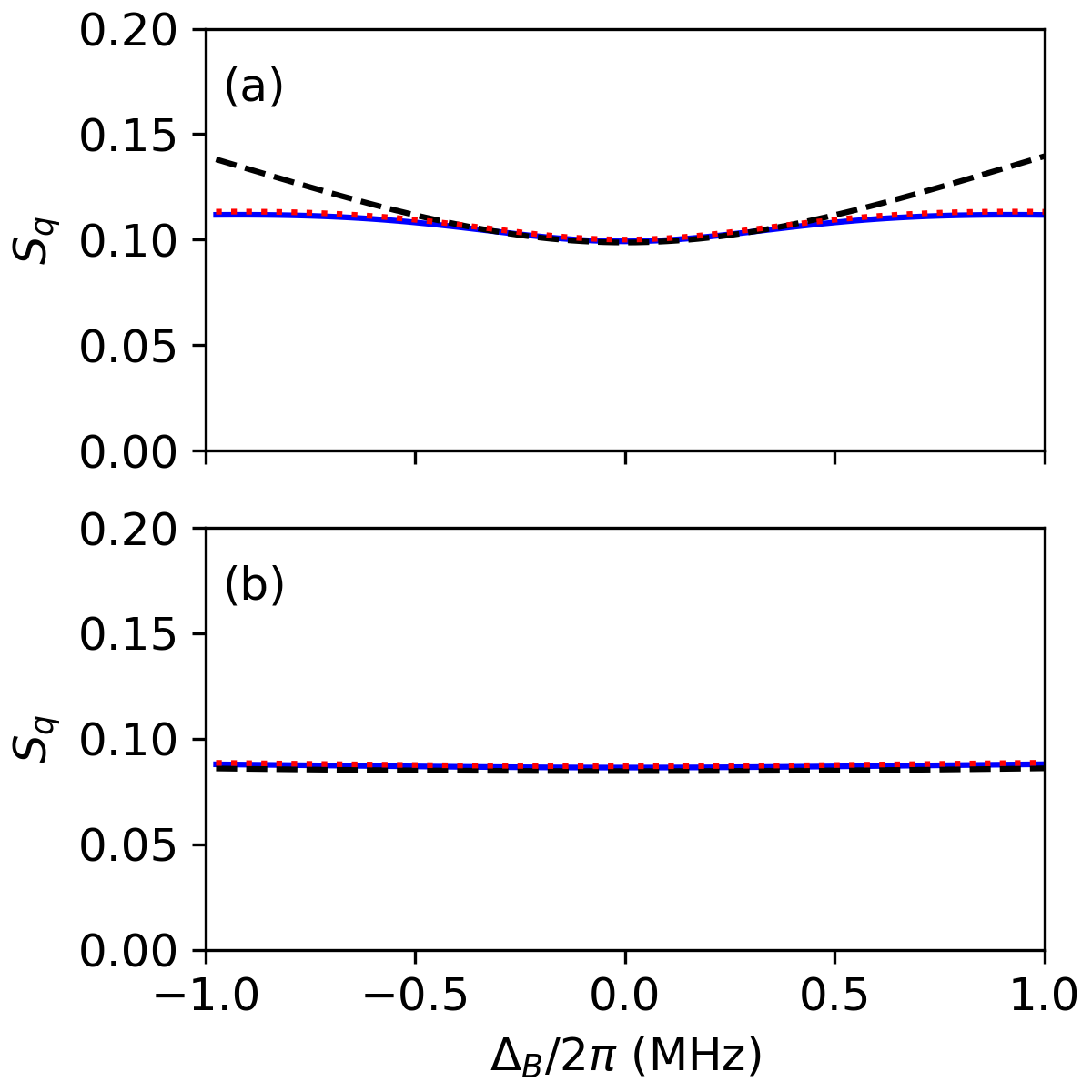}}
\caption{The black dashed, blue solid, and red dotted lines show the synchronization $S_q$ as a function of the Zeeman splitting $\Delta_B$ for the full $(3+3)$-level system, the effective 3-level model, and the perturbative treatment of the effective 3-level model, respectively, for (a) $|\Omega'|=2\pi \times 3.0$~MHz and (b) $|\Omega'|=2\pi \times 6.0$~MHz. The other parameters are 
$\phi_{\pm1}=\phi_0=\phi'=0$, $|\Omega_{\pm1}|=2\pi \times 9.5$~MHz, and $|\Omega_{0}|=2\pi \times 1.0$~MHz (the decay rates $\Gamma_{\text{aux}}'$ and $\Gamma_{\text{aux}}''$ are those for $^{87}$Rb; see Appendix~\ref{sec_appendix0}). }
\label{rewrite_fig_DeltaB}
\end{figure}

Figure~\ref{rewrite_fig_DeltaB} shows  $S_q$ as a function of $\Delta_B$ for two different $\Omega'$, namely $|\Omega'|=2\pi \times 3.0$~MHz and  $|\Omega'|=2\pi \times 6.0$~MHz,  but otherwise the same Rabi coupling strengths and phases as those used in Fig.~\ref{rewrite_fig_alpha}. The black dashed, blue solid, and red dotted lines are for the full $(3+3)$-level system, the effective 3-level model, and the perturbative treatment of the effective 3-level model, respectively. Overall, the agreement between the three different descriptions is very good, except for $|\Delta_B| \gtrsim \pi$~MHz in Fig.~\ref{rewrite_fig_DeltaB}(a).  
It can be seen that $S_q$ depends quite weakly on $\Delta_B$, especially when $|\Omega'/\Delta_B| \gg 1$. 
The weak dependence on $\Delta_B$ follows, for the parameter combinations considered (see also Table~\ref{tab:deltaB_parameters}), from the fact that $S_{q}$ is dominated by the second term in the second pair of absolute value signs in Eq.~(\ref{eq_sync_pt_finite}) and that the decay rates $\Gamma_{\text{control}}$, $\Gamma_{\text{decay}}$, and $\Gamma_{\text{probe}}$ depend comparatively weakly on $\Delta_B$.

In what follows, we focus on $\Delta_B=0$.
Since all elements of the effective spin-1 Hamiltonian are equal to zero for $\Delta_B=0$,  
Fig.~\ref{rewrite_fig_DeltaB} tells us that there exist parameter combinations for which the effective spin-1 system synchronizes
even though the  
 coherent effective coupling $\Omega_{\text{eff}}$ is zero. This remarkable result, namely synchronization due to
purely dissipative effective coupling as opposed to coherent effective coupling, can be traced back to the ``unusual'' functional form of the Lindbladian given in Eq.~(\ref{eq_lindblad_k_4_eff}). The result holds as long as the Rabi coupling strength $\Omega'$  of the decay beams is finite. When $\Omega'$ is equal to zero, the effective master equation does not support a limit cycle state, i.e., the steady-state solution does depend on the initial state (see Appendix~\ref{sec_appendixC}). This result, namely the absence of a limit cycle state for $\Omega'=0$, might be somewhat counterintuitive at first sight since one might expect that the control beams would push population into the $|2\rangle \langle 2|$ state. While this is, indeed, the case, the key point here is that the final state is, as shown in Appendix~\ref{sec_appendixC}, not independent of the initial state.  Correspondingly, phase synchronization---linked to a limit cycle state---only exists for $\Delta_B=0$ when $\Omega'$ is finite but not when $\Omega'$ is zero.

\begin{widetext}
 
\begin{figure}[t]
{\includegraphics[scale=.6]{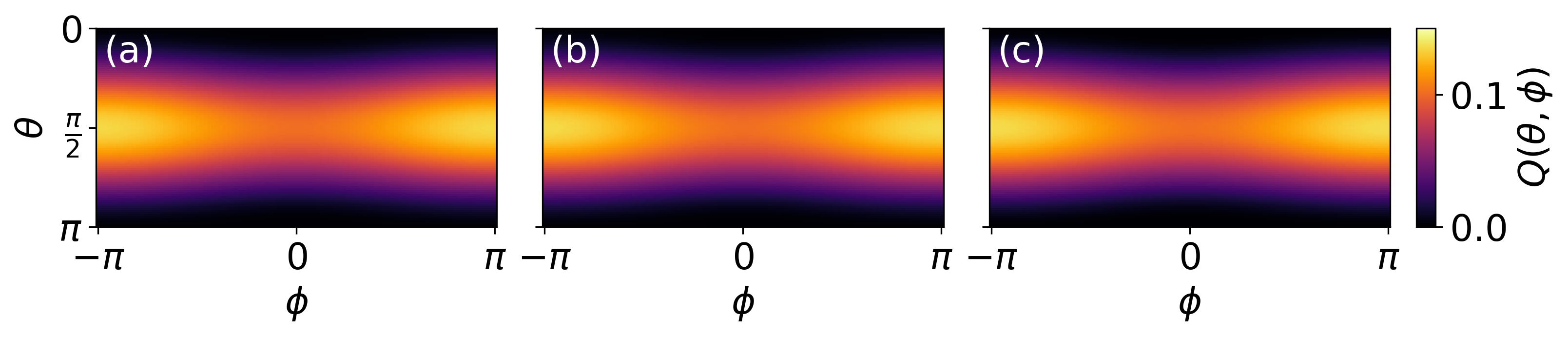}}
\caption{Steady-state Husimi-$Q$ distributions for $\Delta_B=0$. Panels (a), (b), and (c) show results for the full $(3+3)$-level system, the effective 3-level system, and the perturbative treatment of the 3-level system, respectively, for  $\alpha=\phi_{0}=\phi_{-1}=0$, $|\Omega_{+1}|=2\pi \times 9.5$~MHz, $|\Omega_{0}|=2\pi \times 1.0$~MHz, and $|\Omega'|=2\pi \times 3.0$~MHz (the decay rates $\Gamma_{\text{aux}}'$ and $\Gamma_{\text{aux}}''$ are those for $^{87}$Rb; see Appendix~\ref{sec_appendix0}). }
\label{rewrite_fig_HusimiQ}
\end{figure}

\end{widetext}

Since the effective 3-level model does not possess a coherent effective external drive, it is natural to ask what the effective 3-level system synchronizes to when $\Delta_B$ is zero and $\Omega'$ is finite. To shed light on this question, Fig.~\ref{rewrite_fig_HusimiQ} shows the Husimi-$Q$ function for $\Delta_B=0$, calculated  (a) by solving the master equation for the full (3+3)-model, 
(b) by solving the master equation for the effective 3-level model, and
(c) by treating the effective 3-level model within first-order perturbation theory [the Husimi-$Q$ functions are generated using the same Rabi coupling strengths and phases as those used in Fig.~\ref{rewrite_fig_DeltaB}(a)].
The three descriptions agree very well and show  that the Husimi-$Q$ function possesses---consistent with the finite synchronization displayed in Fig.~\ref{rewrite_fig_DeltaB} for $\Delta_B=0$---a maximum at $(\theta,\phi) \approx (\pi/2,\pm \pi)$.

 For $\Delta_B=0$ and $\Omega' \ne 0$, the 
 first-order perturbation theory expression, Eq.~(\ref{eq_sync_pt_finite}),  reduces to
 \begin{eqnarray}
     S_q= 3 \left|
     \cos \left( \frac{\alpha}{2} \right) \right|
     \frac{\Gamma_{\text{control}}^{1/2} \Gamma_{\text{probe}}^{1/2}}{\Gamma_{\text{decay}}+3 \Gamma_{\text{control}}}.
 \end{eqnarray}
 The dependence on $\Gamma_{\text{probe}}$ highlights that the synchronization  arises from a competition between dissipative effective decay into and dissipative effective decay out of the limit cycle state $|2\rangle \langle 2|$.  
 Since the control beam-mediated dissipative processes are much stronger than the decay beam-mediated processes for the parameters used in Figs.~\ref{rewrite_fig_HusimiQ}, we can write
  \begin{eqnarray}
     S_q \approx  \left|
     \cos \left( \frac{\alpha}{2} \right) \right|
     \left(\frac{\Gamma_{\text{probe}}}{ \Gamma_{\text{control}}}\right)^{1/2}.
 \end{eqnarray}
 This expression tells us that an increase of the dissipative processes that correspond to ``self-dephasing'' of state $|2\rangle $ and to dissipative decay from  state $|2\rangle$ to states $|1\rangle$ and $|3\rangle$ (these processes are controlled by $\Gamma_{\text{probe}}$) enhances synchronization while an increase of the dissipative processes that 
 correspond to ``self-dephasing'' of states $|1\rangle $ and $|3\rangle$ and to dissipative decay from state $|1\rangle$ to states $|2\rangle$ and $|3\rangle$ and from state $|2\rangle$ to states $|2\rangle$ and $|1\rangle$ (these processes are controlled by $\Gamma_{\text{control}}$) suppresses synchronization. The fact that the latter  processes includes dissipative decay into state $|2\rangle$ indicates that the mechanism that drives synchronization in the effective spin-1 system that possesses a more involved dissipator structure is fundamentally  different from the mechanism that drives synchronization in the extended ideal 3-level model. 
 In the extended ideal 3-level model, the role of the dissipative terms is to  preserve the limit cycle while the role of the coherent coupling is to synchronize the system. In the effective 3-level system, in contrast, the dissipative terms take on both the roles of  preserving the limit cycle and of synchronizing the system.

 Last, we discuss what happens when we introduce the {\em{ad hoc}} scaling parameter $\beta$. Figure~\ref{rewrite_fig_beta}(e) shows the quantities $S_{q,a1}$ and $S_{q,a2}$ for $\Delta_B=0$. For the parameters chosen (see the figure caption for details), the behavior of $S_{q,a1}$ and $S_{q,a2}$ for $\Delta_B=0$ [Fig.~\ref{rewrite_fig_beta}(e)] is  similar to that for $\Delta_B \ne 0$ [Fig.~\ref{rewrite_fig_beta}(d)] when $\beta$ is close to one but differs when $\beta \lesssim 0.5$. Specifically, both $S_{q,a1}$ and $S_{q,a2}$ go to zero in the $\beta \rightarrow 0$ limit for $\Delta_B=0$, indicating that the finite synchronization for $\beta>0$ originates in the effective incoherent terms.

\section{Conclusions}
\label{sec_outlook}

While classical synchronization has been studied in a wide range of contexts in the social sciences, chemistry, biology, and physics, quantum synchronization has been studied comparatively little. Quantum synchronization---a relative of cooperativity, coherence, and entanglement---quantifies the collective behaviors of a quantum system. This work focused on quantum phase synchronization of a single effective spin-1 system coupled to an environment. The main focus of our study was to explore the role of auxiliary states, which are introduced to realize effective couplings in the ground state manifold. Particular emphasis was placed on quantifying the role of the laser phases. It was pointed out that the  effective incoherent decay into and out of the limit cycle state both play a critical role in understanding the steady-state behaviors of the effective spin-1 system.
While the effective spin-1 system was analyzed through the lense of quantum synchronization, the insights gained are expected to also be instructive for other phenomena that hinge critically on effective dissipative pathways such as, e.g., dissipation-driven quantum phase transitions and dissipation-engineering of initial states.

\section{Acknowledgments}
This work is supported by the W. M. Keck Foundation. Y.Y. acknowledges support through the 
Hong Kong RGC Early Career Scheme (Grant No. 24308323).

\appendix 


\section{Rubidium energy level structure and Hamiltonian in rotating frame}
\label{sec_appendix0}

While the conclusions of this paper apply more generally, the parameters in the main text are applicable to the $^{87}$Rb atom. This appendix introduces the energy level structure of the rubidium-87 atom~\cite{steck}.
The spin-1 system is realized using the $F=1$ states of the 5S$_{1/2}$ ground state manifold. An external magnetic field of strength $B$ along the $z$-axis  sets the quantization axis.  
Owing to the negative $g_F$ factor ($g_F=-1/2$), the energy shifts of the $M_F=1$ and $-1$ states are $-\Delta_B$ and
$\Delta_B$, respectively,
where the magnitude of the energy shift, expressed as an angular frequency, is equal to $2 \pi \times B \times 0.70$~MHz/G. 
The $F=1$ ground state manifold is coupled to the $F''=0$ state of the 5P$_{3/2}$ excited state manifold (D2 line)   via dipole transitions:
\begin{itemize}
\item $|F,M_F\rangle=|1,+1\rangle=|1\rangle \leftrightarrow |F'',M_F''\rangle=|0,0\rangle=|4\rangle $: $\Omega_{\sigma^-}''$, $\vec{k}_{\sigma}''$, $\omega_{\sigma}''$;
\item $|F,M_F\rangle=|1,0\rangle=|2\rangle \leftrightarrow  |F'',M_F''\rangle=|0,0\rangle=|4\rangle $: $\Omega_{\pi}''$, $\vec{k}_{\pi}''$, $\omega_{\pi}''$; and
\item $|F,M_F\rangle=|1,-1\rangle=|3\rangle \leftrightarrow  |F'',M_F''\rangle=|0,0\rangle=|4\rangle $: $\Omega_{\sigma^+}''$, $\vec{k}_{\sigma}''$, $\omega_{\sigma}''$.
\end{itemize}
The $F''=0$ state (state $|4\rangle$) has a life time of $\tau_{\text{aux}}''=26.24(4)$~ns, which corresponds
  to a decay rate of $\Gamma_{\text{aux}}''=(\tau_{\text{aux}}'')^{-1}=2 \pi \times 6.065(9)$~MHz.
  State $|4\rangle$
decays with equal probability to states $|1\rangle$, $|2\rangle$, and $|3\rangle$.
In addition,  states $|1\rangle$ and $|3\rangle$ are coupled to the $F'=1$ states of the 5P$_{1/2}$ excited state manifold [D1 line; life time $\tau'=27.70(4)$~ns, corresponding  to a decay rate $\Gamma_{\text{aux}}'$ of $2 \pi \times 5.746(8)$~MHz] via dipole transitions:
 \begin{itemize}
\item $|F,M_F\rangle=|1,+1\rangle=|1\rangle \leftrightarrow |F',M_F'\rangle=|1,+1\rangle=|5\rangle $: $\Omega_{\pi}'$, $\vec{k}_{\pi}'$, $\omega_{\pi}'$; and
\item $|F,M_F\rangle=|1,-1\rangle=|3\rangle \leftrightarrow |F',M_F'\rangle=|1,-1\rangle=|6\rangle $: $\Omega_{\pi}'$, $\vec{k}_{\pi}'$, $\omega_{\pi}'$.
\end{itemize}
 For the $F'=1$ manifold, $g_{F'}$ is equal to $-1/6$; the energy shifts of the $M_{F'}=+1$ and $-1$ states are $-\Delta_B'$ and $\Delta_B'$, respectively, where 
$\Delta_B'=B \times 2 \pi \times 0.23$~MHz/G.
State $|5\rangle$
decays with equal probability to states $|1\rangle$ and $|2\rangle$ while state $|6\rangle$ decays with equal probability to states $|2\rangle$ and $|3\rangle$. 
Above, $\Omega_{\sigma^{\pm}}''$, $\Omega_{\pi}''$, and  $\Omega_{\pi}'$ denote Rabi coupling strengths; $\vec{k}_{\sigma}''$, $\vec{k}_{\pi}''$, and $\vec{k}_{\pi}'$ wave vectors; and $\omega_{\sigma}''$, $\omega_{\pi}''$, and $\omega_{\pi}'$ angular frequencies of the electric fields that are driving the transitions.
Due to the finite lifetime of states $|4\rangle$, $|5\rangle$, and $|6\rangle$, 
the couplings give rise  to coherent and incoherent processes.

It should be noted that states $|5\rangle$ and $|6\rangle$ also decay to  the $F=2$ ground state hyperfine  manifold of $^{87}$Rb. This implies that the $|1\rangle \leftrightarrow |4\rangle$ and  $|3\rangle \leftrightarrow |6\rangle$ couplings effectively transfer population away from the ground state manifold, i.e., in addition to introducing effective dissipation within the ground state manifold (this is what we want), they open loss channels (for our purposes, this is an  unwanted side effect).  To prohibit population loss, a repump beam could be added that would couple the $F=2$ and $F''=2$ manifolds and would thus  pump population from the $F=2$ to the $F''=2$ manifold. Since population in $F''=2$ has a finite probability to decay to $F=1$ states, this would create a closed system. 
To keep the description simple, the repump laser is not modeled in our calculations and decay of states $|5\rangle$ and $|6\rangle$ to states other than $|1\rangle$, $|2\rangle$, and $|3\rangle$ is neglected.

\begin{figure}[h]
\hspace*{-0.1in}
\vspace*{-0.2in}
{\includegraphics[ scale=.35]{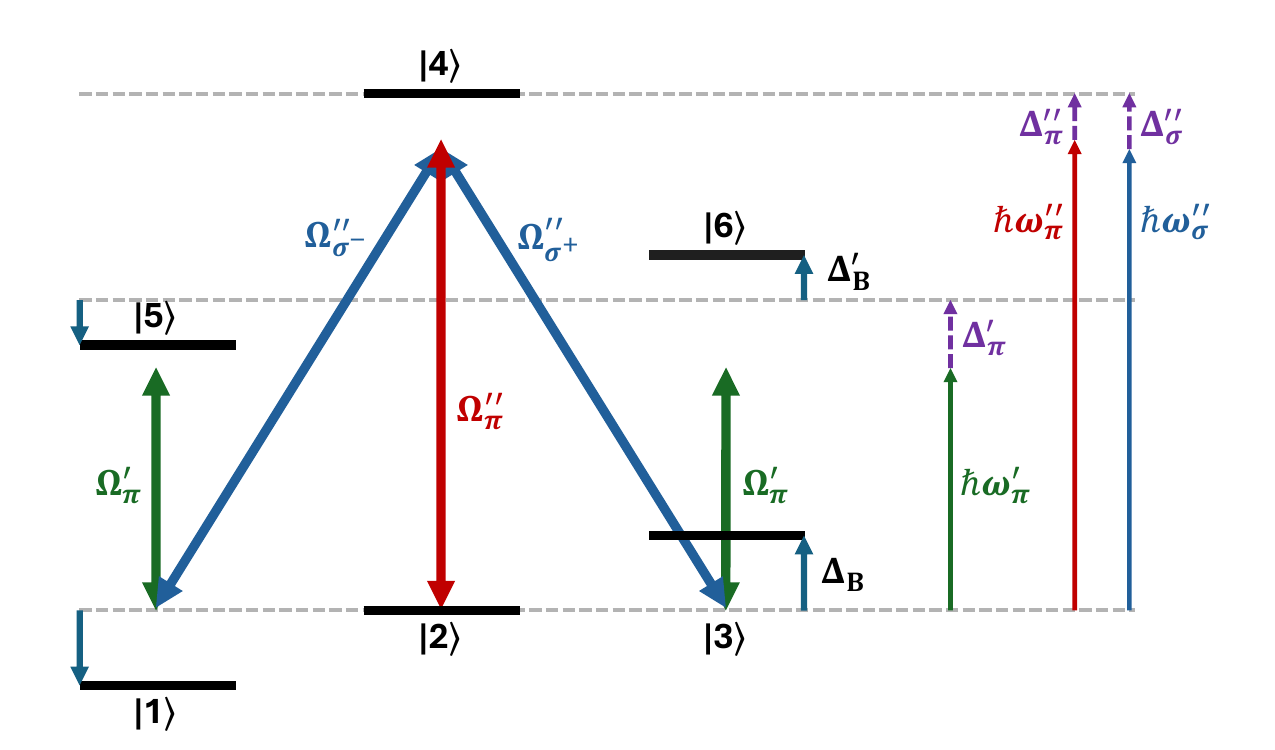}}
\vspace*{0.2in}
\caption{The black horizontal lines show the energy levels. The (3+3)-system consists of the ground states $|1\rangle$, $|2\rangle$, and $|3\rangle$
and the auxiliary states $|4\rangle$, $|5\rangle$, and $|6\rangle$. Gray dashed lines show the energy of the $F=1$ ground state manifold, the $F'=1$ excited state manifold (only the $M_{F'}=\pm1$ states are shown), and the $F''=0$ excited state for vanishing magnetic field. The Zeeman shifts for the $F=1$ ground state manifold and the $F'=1$ excited state manifold are $\Delta_B$ and $\Delta_B'$, respectively. 
Thick arrows indicate coherent couplings between ground and auxiliary states. Thin solid arrows indicate laser frequencies while thin dashed arrows indicate laser detunings.
Level $|1\rangle$ is coupled to $|4\rangle$ with Rabi frequency $\Omega_{\sigma^-}''$, laser frequency $\omega_{\sigma}''$, and detuning $\Delta_{\sigma}''$. Level $|2\rangle$ is coupled to $|4\rangle$ with Rabi frequency $\Omega_{\pi}''$, laser frequency $\omega_{\pi}''$, and detuning $\Delta_{\pi}''$. Level $|3\rangle$ is coupled to $|4\rangle$ with Rabi frequency $\Omega_{\sigma^+}''$, laser frequency $\omega_{\sigma}''$, and detuning $\Delta_{\sigma}''$. Level $|1\rangle$ is coupled to $|5\rangle$ and level $|3\rangle$ is coupled to $|6\rangle$ with Rabi frequency $\Omega_{\pi}'$, laser frequency $\omega_{\pi}'$, and detuning $\Delta_{\pi}'$.}
\label{Rubidium_Level_Scheme_schematic}
\end{figure}

The rubidium-87 atom is modeled by a $(3+3)$-level Hamiltonian, whose Hilbert space is spanned by the states $|j\rangle$ with energies $E_j$ ($j=1-6$), 
\begin{eqnarray}
  \label{eq_e1}
  E_1=E_2-\Delta_B,
\end{eqnarray}
\begin{eqnarray}
    \label{eq_e3}
  E_3=E_2+\Delta_B,
\end{eqnarray}
\begin{eqnarray}
  \label{eq_e7}
  E_4= E_2 +  \omega_{\pi}'' + \Delta_{\pi}'',
\end{eqnarray}
\begin{eqnarray}
  \label{eq_e4}
  E_5=E_2+ \omega_{\pi}' + \Delta_{\pi}' - \Delta_B',
  \end{eqnarray}
  and
\begin{eqnarray}
  \label{eq_e6}
  E_6=E_2 +  \omega_{\pi}' + \Delta_{\pi}' + \Delta_B',
\end{eqnarray}
where $\Delta_{\pi}''$, $\Delta_{\sigma}''$, and $\Delta_{\pi}'$ denote detunings (see Fig.~\ref{Rubidium_Level_Scheme_schematic}).
In writing Eqs.~(\ref{eq_e1})-(\ref{eq_e6}), $E_2$ is used as a ``reference.''
The frequencies $ \omega_{\sigma}''$ and $ \omega_{\pi}''$ 
are related through
\begin{eqnarray}
   \omega_{\sigma}'' +\Delta_{\sigma}''=  \omega_{\pi}'' + \Delta_{\pi}''.
\end{eqnarray}
For simplicity, we assume that the electric field  that drives the transition between states $|1\rangle$ and  $|4\rangle$ can be treated separate from the electric field  that drives the dipole transition between states $|2\rangle$ and  $|4\rangle$, and so on.
 To eliminate the time dependence, which enters through the oscillatory parts of the electric fields [namely, the $\exp(\pm \imath \omega_{\sigma}''t)/2$,  $\exp(\pm \imath \omega_{\pi}'' t)/2$, and $\exp(\pm \imath \omega_{\sigma}' t)/2$ terms] 
 of the Hamiltonian $\hat{H}_{\text{lab}}$, we move to the rotating frame
using the rotation operator $\hat{U}$,
 $\hat{U}= \exp(- \imath \sum_j \omega_j|j\rangle \langle j| t)$,
 where
 $j$ takes the values $j=1-6$; 
 $\omega_j=E_j-\epsilon_j$; and 
\begin{eqnarray}
  \epsilon_1 = -\Delta_B + \Delta_{\pi}'' - \Delta_{\sigma}'',
\end{eqnarray}
\begin{eqnarray}
  \epsilon_2=0,
\end{eqnarray}
\begin{eqnarray}
  \epsilon_3=  \Delta_B+\Delta_{\pi}'' - \Delta_{\sigma}'',
\end{eqnarray}
\begin{eqnarray}
  \epsilon_4=\Delta_{\pi}'' ,
\end{eqnarray}
\begin{eqnarray}
  \epsilon_5=- \Delta_B' -\Delta_{\sigma}''+\Delta_{\pi}''+\Delta_{\pi}',
\end{eqnarray}
and
\begin{eqnarray}
  \epsilon_6=  \Delta_B' -\Delta_{\sigma}''+\Delta_{\pi}''+\Delta_{\pi}'.
\end{eqnarray}

The Hamiltonian $\hat{H}$ 
  in the rotating frame is given by 
 $\hat{U}^{-1}\hat{H}_{\text{lab}} \hat{U}-\imath \hat{U}^{-1} d\hat{U}/dt$. Employing the rotating wave approximation and renaming $\Omega_{\sigma^{\pm}}''=\Omega_{\pm1}$, $\Omega_{\pi}''=\Omega_0$,
 and $\Omega_{\pi}'=\Omega'$, 
we arrive at 
\begin{widetext}
\begin{eqnarray}
    \label{steady_state_eq_hfull_7by7}
    \underline{H} = \frac{1}{2} 
    \left(
    \begin{array}{ccccccc}
2 \epsilon_1
& 0 & 0 &
-|\Omega_{+1}|e^{-i\phi_{+1}} 
& -|\Omega'|e^{-i\phi'} & 0 
\\ 
0 & 0 & 0 &
-|\Omega_{0}|e^{-i \phi_{0}} & 0 & 0 
\\ 
0 & 0 & 2\epsilon_3 &
-|\Omega_{-1}|e^{-i \phi_{-1}} & 0  &
-|\Omega'|e^{-i\phi'}
\\ 
-|\Omega_{+1}|e^{i\phi_{+1}} &
-|\Omega_{0}|e^{i \phi_{0}} &
-|\Omega_{-1}|e^{i \phi_{-1}} &
2\epsilon_4 & 0 & 0 
\\ 
-|\Omega'|e^{i\phi'} & 0 & 0 & 0 &
2 \epsilon_5  & 0 
\\ 
0 & 0 & -|\Omega'|e^{i\phi'} & 0 & 0  &
2 \epsilon_6 
    \end{array}
    \right) .
\end{eqnarray}
\end{widetext}

\section{Master equation reduction}
\label{sec_appendixA}

Our approach employs the formalism developed in Ref.~\cite{reiter2012}. 
In what follows, we work in the rotating frame. 
The Hamiltonian $\hat{H}$
is decomposed into the Hamiltonian $\hat{H}_g$
that lives in the ground state subspace
(spanned by the states $|1\rangle$, $|2\rangle$, and $|3\rangle$),
the Hamiltonian $\hat{H}_{e}$
that lives in the excited state subspace
(spanned by the states $|4\rangle$, $|5\rangle$, and $|6\rangle$),
the excitation operator $\hat{V}_+$
that coherently connects the ground subspace with the excited subspace,
and
the deexcitation operator $\hat{V}_-$
that coherently connects the excited subspace with the ground  subspace
[$(\hat{V}_-)^{\dagger} = \hat{V}_+$],
\begin{eqnarray}
  \hat{H}=\hat{H}_g + \hat{H}_{e} + \hat{V}_+ + \hat{V}_-.
\end{eqnarray}
Formally, the decomposition of the Hamiltonian is accomplished through the
projection operators $\hat{P}_g$ and $\hat{P}_{e}$, which project onto the
ground subspace and excited subspace, respectively,
$\hat{P}_g=\sum_{j=1}^3 |j \rangle \langle j|$ and $\hat{P}_e=\sum_{l=4}^6|l \rangle \langle l|$.  The projection operators obey 
 $ \hat{P}_g + \hat{P}_e = \hat{1}$
and
$  \hat{P}_g  \hat{P}_e = 0$.
With these definitions, one can write  
  $\hat{H}_g = \hat{P}_g \hat{H} \hat{P}_g$,
  $\hat{H}_e = \hat{P}_e \hat{H} \hat{P}_e$,
  $\hat{V}_+ = \hat{P}_e \hat{H} \hat{P}_g$,
and
  $\hat{V}_- = \hat{P}_g \hat{H} \hat{P}_e$.
Note that  $\hat{H}_g$ is diagonal in the  basis that we are working in (this is used in Sec.~\ref{secII_B}).  
The effective Hamiltonian $\hat{{H}}_{\text{eff}}$ reads~\cite{reiter2012}
\begin{eqnarray}
  \label{eq_heff_scheme2}
  \hat{{H}}_{\text{eff}} = \hat{{H}}_g 
   -\frac{1}{2} 
  \sum_{k=1}^3  \left[ \hat{{V}}_- (\hat{{H}}_{\text{NH},k})^{-1}
    \hat{{V}}_{+,k}
+
\mbox{h.c.} \right]
  ,
  \end{eqnarray}
  where ``h.c.'' stands for ``hermitian conjugate'' and where the ``projected 
   non-hermitian Hamiltonians'' $\hat{{H}}_{\text{NH},k}$, 
   \begin{eqnarray}
   \label{eq_ham_nh_proj}
  (\hat{{H}}_{\text{NH},k})^{-1} = \left(  \hat{{H}}_{\text{NH}} - \epsilon_{k}   \right)^{-1},
\end{eqnarray}
are defined in terms of the ``non-hermitian Hamiltonian''
$\hat{{H}}_{\text{NH}}$, 
\begin{eqnarray}
  \hat{{H}}_{\text{NH}}
  =
  \hat{{H}}_e - \frac{\imath}{2 } \Bigg[ \sum_{k=1}^3 \frac{\Gamma_{\text{aux}}''}{3} \hat{{{{L}}}}_{k,4}^{\dagger} \hat{{{{L}}}}_{k,4}
  + \nonumber \\
  \sum_{k=1}^2  \frac{\Gamma_{\text{aux}}'}{2} \hat{{{{L}}}}_{k,5}^{\dagger} \hat{{{{L}}}}_{k,5}
  +
  \sum_{k=2}^3  \frac{\Gamma_{\text{aux}}'}{2} \hat{{{{L}}}}_{k,6}^{\dagger} \hat{{{{L}}}}_{k,6}
  \Bigg]
.
\end{eqnarray}
Note that $\hat{H}_e$ is non-hermitian due to the finite lifetime of states $|j\rangle$, $j=4-6$.
Equation~(\ref{eq_heff_scheme2})
shows that $\hat{{H}}_{\text{eff}}$ is hermitian by construction. The projected excitation operators $\hat{{V}}_{+,k}$
($k=1-3$)  are defined in terms of the projectors $\hat{P}_{k}$, 
  $\hat{{V}}_{+,k} = \hat{{V}}_{+} \hat{P}_{k}$
and
  $\hat{P}_{k}	= |k \rangle \langle k|$.
With these definitions,
the effective Lindbladian $\hat{{{{L}}}}_{\text{eff},k,l}$ can be written as 
\begin{eqnarray}
\label{eq_leff_scheme2}
\hat{{{{L}}}}_{\text{eff},k,l}
  =
  \sum_{j=1}^3
\hat{{{{R}}}}_{k,j}^{(l)},
\end{eqnarray}
where
\begin{eqnarray}
\label{eq_leff_scheme2_final}
  \hat{{{{R}}}}_{k,j}^{(l)}
  =
\sqrt{  \Gamma_{k,l} }\hat{{{{L}}}}_{k,l}  
  (\hat{{H}}_{\text{NH},j})^{-1} \hat{{V}}_{+,j}.
\end{eqnarray}
The quantity $|\langle k|\hat{{{{R}}}}_{k,j}^{(l)}|j\rangle |^2$ is the effective rate for making a dissipative transition from state $|j\rangle$ to state $|k\rangle$. Dissipative transitions into state $|k\rangle$ can occur from states $|1\rangle$, $|2\rangle$, and $|3\rangle$. If $j$ is equal to $k$, the process may be interpreted as an auxiliary state-mediated  ``self-dissipation.'' Importantly,  the ``auxiliary operators'' $\hat{{{{R}}}}_{k,j}^{(l)}$ themselves are not to be interpreted as Lindbladian.

\section{Ideal effective 3-level Hamiltonian}
\label{sec_appendixB}

The master equation reduction yields the effective 3-level Hamiltonian $\underline{H}_{\text{eff}} $,
\begin{widetext}
\begin{eqnarray}
\label{eq_heff_full}
\underline{H}_{\text{eff}} = \nonumber \\
\frac{1}{8}
\left(
\begin{array}{ccc}
h_1
& 
  -\frac{\Omega_{0}\Omega_{+1}^*}{-\frac{i \Gamma_{\text{aux}}''}{2} + \Delta_{\pi}''} - \frac{\Omega_{0} \Omega_{+1}^*}{\frac{i \Gamma_{\text{aux}}''}{2}+\Delta_{\sigma}''+\Delta_B}  
& 
  -\frac{\Omega_{-1}\Omega_{+1}^*}{-\frac{i \Gamma_{\text{aux}}''}{2}+\Delta_{\sigma}''-\Delta_{B}}- \frac{\Omega_{-1}\Omega_{+1}^*}{\frac{i \Gamma_{\text{aux}}''}{2}+\Delta_{\sigma}''+\Delta_{B}}  
\\ 
  -\frac{\Omega_{0}^*\Omega_{+1}}{\frac{i \Gamma_{\text{aux}}''}{2} + \Delta_{\pi}''} - \frac{\Omega_{0}^* \Omega_{+1}}{-\frac{i \Gamma_{\text{aux}}''}{2}+\Delta_{\sigma}''+\Delta_B}
&
 h_2
&
 -\frac{\Omega_{-1} \Omega_{0}^*}{-\frac{i \Gamma_{\text{aux}}''}{2}+\Delta_{\sigma}''-\Delta_{B}}- \frac{\Omega_{-1} \Omega_{0}^*}{\frac{i \Gamma_{\text{aux}}''}{2}+\Delta_{\pi}''}  
\\ 
  -\frac{\Omega_{-1}^*\Omega_{+1}}{\frac{i \Gamma_{\text{aux}}''}{2}+\Delta_{\sigma}''-\Delta_{B}}- \frac{\Omega_{-1}^*\Omega_{+1}}{-\frac{i \Gamma_{\text{aux}}''}{2}+\Delta_{\sigma}''+\Delta_{B}}   
&
 -\frac{\Omega_{-1}^*\Omega_{0}}{\frac{i \Gamma_{\text{aux}}''}{2}+\Delta_{\sigma}''-\Delta_{B}}- \frac{\Omega_{-1}^* \Omega_{0}}{-\frac{i \Gamma_{\text{aux}}''}{2}+\Delta_{\pi}''}   
& 
h_3
\end{array}
\right),
\end{eqnarray}
where the scaled effective Stark shifts $h_1$, $h_2$, and $h_3$ are given by
\begin{eqnarray}
h_1=
8(\Delta_{\pi}''-\Delta_{\sigma}''-\Delta_B)- \left[\left( \frac{|\Omega_{+1}|^2}{-\frac{i \Gamma_{\text{aux}}''}{2} +\Delta_{\sigma}''+\Delta_B} + \frac{|\Omega'|^2}{-\frac{i \Gamma_{\text{aux}}'}{2} +\Delta_{\pi}'-\Delta_B' +\Delta_B}\right)+\text{c.c.} \right] ,
\end{eqnarray}
\begin{eqnarray}
h_2= -\frac{|\Omega_{0}|^2}{-\frac{i \Gamma_{\text{aux}}''}{2}+\Delta_{\pi}''}+\text{c.c.} ,
\end{eqnarray}
and
\begin{eqnarray}
h_3=8(\Delta_{\pi}''-\Delta_{\sigma}''+\Delta_B)- \left[\left( \frac{|\Omega_{-1}|^2}{-\frac{i \Gamma_{\text{aux}}''}{2} +\Delta_{\sigma}''-\Delta_B} + \frac{|\Omega'|^2}{-\frac{i \Gamma_{\text{aux}}'}{2} +\Delta_{\pi}'+\Delta_B' -\Delta_B}\right)+\text{c.c.} \right] .
\end{eqnarray}
\end{widetext}
To map this Hamiltonian onto the ideal spin-1 Hamiltonian $\underline{H}_S$, we must demand
equal spacing and the absence of coherent coupling between states $|1\rangle$ and $|3\rangle$, i.e., we must demand $h_1-h_2=h_2-h_3$ and $H_{\text{eff},1,3}=0$.
These conditions are fulfilled for $\Delta_{\pi}''=\Delta_{\sigma}''=\Delta_{\pi}'=0$.
 In this case, Eq.~(\ref{eq_heff_full}) reduces to Eqs.~(9)-(13) from the main text.

Table~\ref{tab:deltaB_parameters} shows the parameters of the effective 3-level system for the parameters employed in Figs.~\ref{rewrite_fig_alpha}-\ref{rewrite_fig_HusimiQ}.

\begin{widetext}

\begin{table}[h!]
\centering
\begin{tabular}{l|ccc}
 & $\Delta_B = 0$ & $\Delta_B = 2\pi \times 0.2$ MHz & $\Delta_B = 2\pi \times 0.4$ MHz \\
\hline
$|\Delta_{\text{eff}}|$ & 0 & $2\pi \times 0.725$~MHz & $2\pi \times 1.437$~MHz \\
$|H_{\text{eff},2,3}|$ & 0 & $2\pi \times 0.026$~MHz & $2\pi \times 0.051$~MHz  \\
$\Gamma_{\text{control}}$ & $2\pi \times 4.961$~MHz & $2\pi \times 4.939$~MHz & $2\pi \times 4.875$~MHz \\
$\Gamma_{\text{probe}}$ & $2\pi \times 0.055$~MHz & $2\pi \times 0.055$~MHz & $2\pi \times 0.055$~MHz \\
$\Gamma_{\text{decay}}$ & $2\pi \times 0.783$~MHz & $2\pi \times 0.781$~MHz & $2\pi \times 0.776$~MHz \\
$\Gamma_{\text{control}}^{1/2}\Gamma_{\text{probe}}^{1/2}$ & $2\pi \times 0.522$~MHz & $2\pi \times 0.521$~MHz & $2\pi \times 0.518$~MHz 
\end{tabular}
\caption{Parameters of the effective 3-level system for three different $\Delta_B$ and $|\Omega_{\pm1}|=2\pi \times 9.5$~MHz, $|\Omega_{0}|=2\pi \times 1.0$~MHz, and  $|\Omega'|=2\pi \times 3.0$~MHz (the decay rates $\Gamma_{\text{aux}}'$ and $\Gamma_{\text{aux}}''$ are those for $^{87}$Rb).}
\label{tab:deltaB_parameters}
\end{table}

\end{widetext}

\section{Perturbation theory for the ``full'' $(3+3)$-level master equation}
\label{sec_appendixBpert}

This appendix describes a perturbative approach for determining the stationary solution of the master equation in the rotating frame, assuming that the Hamiltonian in the rotating frame is time independent~\cite{koppenhofer2019,li2014}.
The Hamiltonian $\hat{H}$
is divided into two pieces,
$\hat{H}=\hat{H}_{\text{ref}}+ \lambda \hat{H}_{\text{pert}}$, where $\hat{H}_{\text{ref}}$
is chosen such that its stationary solution is known and $\lambda$ denotes a parameter that facilitates keeping track of orders (at the end, $\lambda$ will be set to $1$)~\cite{koppenhofer2019,li2014}. 
In our case, $\hat{H}_{\text{ref}}$ can be conveniently defined as $\lim_{|\Omega_0| \rightarrow 0} \hat{H}$
and $\hat{H}_{\text{pert}}$ as $\hat{H}-(\lim_{|\Omega_0| \rightarrow 0} \hat{H})$. 
With these definitions, the master equation
can be written as
\begin{eqnarray}
\label{eq_masterequation_full}
    \frac{d \hat{\rho}}{dt}=
    \hat{\cal{L}}_{\text{ref}} \hat{\rho} 
    +\lambda \hat{\cal{L}}_{\text{pert}} \hat{\rho},
\end{eqnarray}
where
\begin{eqnarray}
    \hat{\cal{L}}_{\text{ref}} \hat{\rho} =
    -\imath [ \hat{H}_{\text{ref}}, \hat{\rho}]+
    \sum_j \gamma_j \hat{\cal{D}}[\hat{{L}}_j](\hat{\rho})
\end{eqnarray}
and
\begin{eqnarray}
    \hat{\cal{L}}_{\text{pert}} \hat{\rho} =
    -\imath [\hat{H}_{\text{pert}},\hat{\rho}].
\end{eqnarray}
Note that we denote the Lindbladian by $\hat{L}_j$ with associated decay rate $\gamma_j$ (no double indices as in the main text) throughout this appendix to simplify the notation.
Making the ansatz 
\begin{eqnarray}
\label{eq_ansatz}
    \hat{\rho}= \frac{1}{{\cal{N}}}\sum_{k=0}^{\infty}
\lambda^k \hat{\rho}^{(k)},
\end{eqnarray}
 where ${\cal{N}}$ denotes a normalization constant, and defining $\hat{\rho}^{(0)}$ through
$\hat{\cal{L}}_{\text{ref}} \hat{\rho}^{(0)}=0$, 
the higher-order contributions $\hat{\rho}^{(k)}$ with $k=1,2,\cdots$ are determined by
\begin{eqnarray}
\label{eq_recursion_simple}
    \hat{\cal{L}}_{\text{ref}} \hat{\rho}^{(k)} = 
    - \hat{\cal{L}}_{\text{pert}} \hat{\rho}^{(k-1)}.
\end{eqnarray}
This recursive relationship for the stationary solution is found by inserting the ansatz given in Eq.~(\ref{eq_ansatz}) into Eq.~(\ref{eq_masterequation_full}), setting the time derivatives to zero, 
and grouping terms with the same power of $\lambda$. 
Equation~(\ref{eq_recursion_simple}) defines a set of linear equations that can be solved under the constraint $\text{Tr}(\hat{\rho}^{(0)})=1$.

The perturbative framework assumes that $\hat{H}_{\text{pert}}$ can be treated as a perturbation that does not ``deform'' the stationary reference solution $\hat{\rho}^{(0)}$ too much, i.e., when the  reference density matrix $\hat{\rho}^{(0)}$ is a limit cycle state and the drive strength is sufficiently small.  
Since the perturbative framework does not  ensure positivity~\cite{li2014}, it cannot---in general---be used to calculate observables that depend on the eigenvalues of the density matrix, such as the von Neumann entropy,  or that can be formulated in terms of the purity of the density. For the ideal spin-1 system, one can show that one of the eigenvalues of the perturbatively determined density matrix is negative at all orders of perturbation theory. 

It should be noted that the perturbation theory formalism introduced in this section assumes---when grouping terms of equal order in $\lambda$---that the dissipators are  independent of $\lambda$, i.e., independent of the drive strength. This is the case for the ``full'' master equation but not for the effective master equation, Eq.~(\ref{steady_states_eq_me_full}). Appendix~\ref{sec_appendixD} develops a perturbative framework for obtaining  stationary solutions to Eq.~(\ref{steady_states_eq_me_full}).

\section{Perturbation theory for the effective 3-level master equation}
\label{sec_appendixD}

The perturbation theory framework for finding the stationary solution to the effective master equation, such as Eq.~(\ref{steady_states_eq_me_full}), starts by identifying a reference Hamiltonian $\hat{H}_{\text{eff,ref}}$ and reference dissipators such that the reference master equation has a limit cycle solution, which can be found analytically or numerically.
Formally, we write
\begin{eqnarray}
\label{eq_ham_appendixc1}
    \hat{H}_{\text{eff}}=
    \hat{H}_{\text{eff,ref}}+\lambda
    \hat{H}_{\text{eff,pert}}
\end{eqnarray}
and
\begin{eqnarray}
    \hat{{L}}_{\text{eff},j}=
    \hat{{L}}_{\text{eff,ref},j}+\lambda
    \hat{{L}}_{\text{eff,pert},j} 
\end{eqnarray}
such that
\begin{eqnarray}
    \hat{\cal{L}}_{\text{eff,ref}} \hat{\rho}_{\text{eff}} = 0,
\end{eqnarray}
where
\begin{eqnarray}
    \hat{\cal{L}}_{\text{eff,ref}} \hat{\rho}_{\text{eff}} =
    -\imath [ \hat{H}_{\text{eff,ref}},\hat{\rho}_{\text{eff}}] + \sum_j {\hat{\cal{D}}}[\hat{{L}}_{\text{eff,ref},j}](\hat{\rho}_{\text{eff}}).
\end{eqnarray}
Note that we denote the effective Lindbladian by $\hat{L}_{\text{eff},j}$  (no double indices as in the main text) throughout this appendix to simplify the notation.
In Eq.~(\ref{eq_ham_appendixc1}), we assume that $\hat{H}_{\text{eff}}$ does not contain any quadratic terms in $\lambda^2$ (this holds for the effective $3 \times 3$ Hamiltonian with $\Delta_{\pi}''=\Delta_{\sigma}''=\Delta_{\pi}'=0$). As in the previous section, $\lambda$ will be set equal to one at the end.
Defining 
\begin{eqnarray}
\label{eq_lastappendix}
\hat{\cal{L}}_{\text{eff,pert},1} \hat{\rho}_{\text{eff}}
    =
    -\imath [ \hat{H}_{\text{eff,pert}},\hat{\rho}_{\text{eff}}]
+\nonumber \\
    \sum_j \hat{\cal{C}}[\hat{{L}}_{\text{eff,ref},j}, \hat{{L}}_{\text{eff,pert},j}](\hat{\rho}),
\end{eqnarray}
\begin{eqnarray}
    \hat{\cal{C}}[\hat{{L}}_{\text{eff,ref},j}, \hat{{L}}_{\text{eff,pert},j}](\hat{\rho}_{\text{ref}})
    = \nonumber \\ \hat{{L}}_{\text{eff,ref},j} 
    \hat{\rho}_{\text{eff}}
    (\hat{{L}}_{\text{eff,pert},j}  )^{\dagger}+\nonumber \\
     \hat{{L}}_{\text{eff,pert},j} 
    \hat{\rho}_{\text{eff}}
    (\hat{{L}}_{\text{eff,ref},j}  )^{\dagger}  -
    \nonumber \\
     \frac{1}{2} \left\{
     (\hat{{L}}_{\text{eff,ref},j}  )^{\dagger} \hat{{L}}_{\text{eff,pert},j}, 
    \hat{\rho}_{\text{eff}}
    \right\}
    - \nonumber \\
    \frac{1}{2}  \left\{
     (\hat{{L}}_{\text{eff,pert},j}  )^{\dagger} \hat{{L}}_{\text{eff,ref},j},
     \hat{\rho}_{\text{eff}}
    \right\}
    ,
\end{eqnarray}
and
\begin{eqnarray}
    \hat{\cal{L}}_{\text{eff,pert},2}\hat{\rho}_{\text{eff}} =\sum_{j} \hat{\cal{D}}[ \hat{{L}}_{\text{eff,pert},j}](\hat{\rho}_{\text{eff}})
    ,
\end{eqnarray}
the effective master equation reads
\begin{eqnarray}
    \dot{\hat{\rho}}_{\text{eff}}=
    \hat{\cal{L}}_{\text{eff,ref}} \hat{\rho}_{\text{eff}}
    +
    \lambda  \hat{\cal{L}}_{\text{eff,pert},1} \hat{\rho}_{\text{eff}} 
    + \lambda^2 
    \hat{\cal{L}}_{\text{eff,pert},2} 
    \hat{\rho}_{\text{eff}}.\nonumber \\
\end{eqnarray}
Making the ansatz
\begin{eqnarray}
    \hat{\rho}_{\text{eff}}= \frac{1}{\cal{N}}
    \sum_{k=0}^{\infty} \lambda^k \hat{\rho}_{\text{eff}}^{(k)},
\end{eqnarray}
the stationary  solution can be obtained recursively order by order: 
\begin{eqnarray}
    \hat{\cal{L}}_{\text{eff,ref}} \hat{\rho}_{\text{eff}}^{(0)}
    =0,
\end{eqnarray}
\begin{eqnarray}
    \hat{\cal{L}}_{\text{eff,ref}} \hat{\rho}_{\text{eff}}^{(1)}+
    \hat{\cal{L}}_{\text{eff,pert},1} \hat{\rho}_{\text{eff}}^{(0)}
    =0,
\end{eqnarray}
and, for $k \ge 2$,
\begin{eqnarray}
    \hat{\cal{L}}_{\text{eff,ref}} \hat{\rho}_{\text{eff}}^{(k)}+
    \hat{\cal{L}}_{\text{eff,pert},1} \hat{\rho}_{\text{eff}}^{(k-1)}  
+
\hat{\cal{L}}_{\text{eff,pert},2}\hat{\rho}_{\text{eff}}^{(k-2)}  =0.
\nonumber \\
\end{eqnarray}
Compared to the formalism outlined in Appendix~\ref{sec_appendixBpert}, the formulation for the effective master equation contains  additional terms, namely  the second term on the right hand side of Eq.~(\ref{eq_lastappendix}) as well as 
the superoperator $\hat{\cal{L}}_{{\text{eff,pert},2}}$. These terms arise since the Lindbladians 
$\hat{{L}}_{\text{eff},j}$ contain terms that are proportional to the quantity that serves as the small parameter. Since the effective Lindbladian can contain several terms [see Eq.~(\ref{eq_leff_scheme2})], ``isolating'' a reference Lindbladian introduces first-order cross terms as well as second-order terms. 

For the effective Hamiltonian $\hat{H}_{\text{eff}}$ [Eq.~(\ref{eq_eff_H_ideal})], 
$\hat{H}_{\text{eff,ref}}$ is given by 
the diagonal Hamiltonian $(\hat{H}_{\text{eff}})|_{|\Omega_0|=0}$ while $\hat{H}_{\text{eff,pert}}$ is defined through
$\hat{H}_{\text{eff,pert}}=\hat{H}_{\text{eff}}-\hat{H}_{\text{eff,ref}}$, i.e., $\hat{H}_{\text{eff,pert}}$ contains non-zero off-diagonal elements that are proportional to $|\Omega_0|$.
The effective reference Lindbladian $\hat{{L}}_{\text{eff,ref},j,k}$ 
are given by $(\hat{{L}}_{\text{eff},j,k})|_{|\Omega_0|=0}$
and the Lindbladian $\hat{{L}}_{\text{eff,pert},j,k}$ are defined through 
$\hat{{L}}_{\text{eff,pert},j,k}=\hat{{L}}_{\text{eff},j,k}-\hat{{L}}_{\text{eff,ref},j,k}$.

\section{Limit cycle existence}
\label{sec_appendixC}
This appendix analyzes the time-dependent and steady-state solutions for $\Delta_{\pi}''=\Delta_{\sigma}''=\Delta_{\pi}'=0$ and $|\Omega_{+1}|=|\Omega_{-1}|$ (we denote this Rabi coupling by $|\Omega_c|$)
in the absence of the external drive, i.e., for $\Omega_0=0$.
 An analysis of the steady-state solution is required since  the existence of a limit cycle state is a prerequisite for synchronization.
We provide analytical solutions for the time-dependent density matrix elements for three cases:
\begin{itemize} 
\item Case (i): $\Delta_B=0$, $\phi_{\pm 1}$ finite, and $\Omega'=0$ (limit cycle state is not supported).
\item Case (ii): $\Delta_B=0$, $\phi_{\pm 1}=0$,  and $\Omega'$ finite (limit cycle state $|2\rangle \langle 2|$).
\item Case (iii): $\Delta_B$ finite, $\Omega_{\pm1}=0$, and $\Omega'$ finite (limit cycle state $|2\rangle \langle 2|$).
\end{itemize}
While we do not provide analytical solutions for finite $\Delta_B$ and arbitrary $\Omega_{\pm1}$ and $\Omega'$, it can be readily shown that  the state $|2\rangle \langle 2|$ is a limit-cycle, provided either $|\Omega_{\pm1}|$ or
$|\Omega'|$ are finite.
We emphasize the distinct characteristics for vanishing and finite $\Delta_B$: The system with $\Omega'=0$ and finite $\Omega_{\pm 1}$  does not support a limit cycle state for $\Delta_B=0$ but does support a limit cycle state, namely the state $|2\rangle \langle 2|$, for $\Delta _B \ne 0$.

\subsection{Case (i): $\Delta_B=0$, $\phi_{\pm 1}$ finite, and $\Omega'=0$}
The effective 3-level master equation reads
\begin{widetext}
\begin{eqnarray}
\dot{\rho}_{\text{eff},1,1}(t) =
\Gamma_{\text{eff}}''\left[
-\rho_{\text{eff},1,1}(t)-\frac{1}{4}e^{-i(\phi_{-1}-\phi_{+1})}\rho_{\text{eff},1,3}(t)-\frac{1}{4}e^{i(\phi_{-1}-\phi_{+1})}\rho_{\text{eff},3,1}(t)+\frac{1}{2}\rho_{\text{eff},3,3}(t)
\right],
\end{eqnarray}
\begin{eqnarray}
\dot{\rho}_{\text{eff},1,2}(t) =
\Gamma_{\text{eff}}''\left[
-\frac{3}{4}\rho_{\text{eff},1,2}(t)-\frac{3}{4}e^{i(\phi_{-1}-\phi_{+1})}\rho_{\text{eff},3,2}(t) \right],
\end{eqnarray}
\begin{eqnarray}
\dot{\rho}_{\text{eff},1,3}(t) =
\Gamma_{\text{eff}}''\left[
-\frac{3}{4}e^{i(\phi_{-1}-\phi_{+1})}\rho_{1,1}(t)-\frac{3}{2}\rho_{\text{eff},1,3}(t)-\frac{3}{4}e^{i(\phi_{-1}-\phi_{+1})}\rho_{\text{eff},3,3}(t) \right],
\end{eqnarray}
\begin{eqnarray}
\dot{\rho}_{\text{eff},2,2} =
\Gamma_{\text{eff}}''\bigg[
\frac{1}{2}\rho_{\text{eff},1,1}(t)+\frac{1}{2}e^{-i(\phi_{-1}-\phi_{+1})}\rho_{\text{eff},1,3}(t)+\frac{1}{2}e^{i(\phi_{-1}-\phi_{+1})}\rho_{3,1}(t)+ 
\frac{1}{2}\rho_{\text{eff},3,3}(t) \bigg],
\end{eqnarray}
\begin{eqnarray}
\dot{\rho}_{\text{eff},2,3}(t) =
\Gamma_{\text{eff}}''\left[
-\frac{3}{4}e^{i(\phi_{-1}-\phi_{+1})}\rho_{\text{eff},2,1}(t)-\frac{3}{4}\rho_{\text{eff},2,3}(t)
\right],
\end{eqnarray}
and
\begin{equation}
\dot{\rho}_{\text{eff},3,3}(t) =
\Gamma_{\text{eff}}''\left[
\frac{1}{2}\rho_{1,1}(t)-\frac{1}{4}e^{i(\phi_{-1}-\phi_{+1})}\rho_{\text{eff},1,3}(t)-\frac{1}{4}e^{-i(\phi_{-1}-\phi_{+1})}\rho_{3,1}(t)-\rho_{\text{eff},3,3}(t)
\right],
\end{equation}
where the effective decay rate $\Gamma_{\text{eff}}''$ is given by
\begin{equation}
\label{eq_gamma_eff}
    \Gamma_{\text{eff}}''=\frac{2|\Omega_{c}|^2}{3\Gamma_{\text{aux}}''}.
\end{equation}
We note that the diagonal elements $\rho_{\text{eff},k,k}(t)$ and corner elements 
$\rho_{\text{eff},1,3}(t)$ of the differential equation are coupled 
and that the 
first off-diagonal elements $\rho_{\text{eff},1,2}(t)$ and $\rho_{\text{eff},2,3}(t)$
are coupled. The solution for an arbitrary initial state $\underline{\rho}_{\text{eff}}(0)$ reads
\begin{eqnarray}
\rho_{\text{eff},1,1}(t)=\rho_{\text{eff},1,1}(0)\left(\frac{3}{8}+\frac{1}{2}e^{-\frac{3}{2}\Gamma_{\text{eff}}''t}+\frac{1}{8}e^{-2\Gamma_{\text{eff}}''t}\right)+\Re \left(\rho_{\text{eff},1,3}(0)e^{-i(\phi_{-1}-\phi_{+1})} \right)\left(-\frac{1}{4}+\frac{1}{4}e^{-2\Gamma_{\text{eff}}''t}\right)
\nonumber \\
+\rho_{\text{eff},3,3}(0)\left(\frac{3}{8}-\frac{1}{2}e^{-\frac{3}{2}\Gamma_{\text{eff}}''t}+\frac{1}{8}e^{-2\Gamma_{\text{eff}}''t}\right),
\end{eqnarray}
\begin{eqnarray}
\rho_{\text{eff},1,2}(t)=\rho_{\text{eff},1,2}(0)\left(\frac{1}{2}+\frac{1}{2}e^{-\frac{3}{2}\Gamma_{\text{eff}}''t}\right)+\rho_{\text{eff},2,3}^*(0)e^{i(\phi_{-1}-\phi_{+1})}\left(-\frac{1}{2}+\frac{1}{2}e^{-\frac{3}{2}\Gamma_{\text{eff}}''t}\right),
\end{eqnarray}
\begin{eqnarray}
\rho_{\text{eff},1,3}(t)=\rho_{\text{eff},1,1}(0)e^{i(\phi_{-1}-\phi_{+1})}\left(-\frac{3}{8}+\frac{3}{8}e^{-2\Gamma_{\text{eff}}''t}\right)+\rho_{\text{eff},1,3}(0)\left(\frac{1}{8}+\frac{1}{2}e^{-\frac{3}{2}\Gamma_{\text{eff}}''t}+\frac{3}{8}e^{-2\Gamma_{\text{eff}}''t}\right)
\nonumber \\
+\rho_{\text{eff},1,3}^*(0)e^{2i(\phi_{-1}-\phi_{+1})}\left(\frac{1}{8}-\frac{1}{2}e^{-\frac{3}{2}\Gamma_{\text{eff}}''t}+\frac{3}{8}e^{-2\Gamma_{\text{eff}}''t}\right)+\rho_{\text{eff},3,3}(0)e^{i(\phi_{-1}-\phi_{+1})}\left(-\frac{3}{8}+\frac{3}{8}e^{-2\Gamma_{\text{eff}}''t}\right),
\end{eqnarray}
\begin{eqnarray}
\rho_{\text{eff},2,2}(t)=1+\rho_{\text{eff},1,1}(0)\left(-\frac{3}{4}-\frac{1}{4}e^{-2\Gamma_{\text{eff}}''t}\right)+\Re\left(\rho_{\text{eff},1,3}(0)e^{-i(\phi_{-1}-\phi_{+1})}\right)\left(\frac{1}{2}-\frac{1}{2}e^{-2\Gamma_{\text{eff}}''t}\right)+
\nonumber \\
\rho_{\text{eff},3,3}(0)\left(-\frac{3}{4}-\frac{1}{4}e^{-2\Gamma_{\text{eff}}''t}\right),
\end{eqnarray}
\begin{eqnarray}
\rho_{\text{eff},2,3}(t)=\rho_{\text{eff},1,2}^*(0)e^{i(\phi_{-1}-\phi_{+1})}\left(-\frac{1}{2}+\frac{1}{2}e^{-\frac{3}{2}\Gamma_{\text{eff}}''t}\right)+\rho_{\text{eff},2,3}(0)\left(\frac{1}{2}+\frac{1}{2}e^{-\frac{3}{2}\Gamma_{\text{eff}}''t}\right),
\end{eqnarray}
and
\begin{eqnarray}
\rho_{\text{eff},3,3}(t)=\rho_{\text{eff},1,1}(0)\left(\frac{3}{8}-\frac{1}{2}e^{-\frac{3}{2}\Gamma_{\text{eff}}''t}+\frac{1}{8}e^{-2\Gamma_{\text{eff}}''t}\right)+\Re\left(\rho_{\text{eff},1,3}(0)e^{-i(\phi_{-1}-\phi_{+1})}\right)\left(-\frac{1}{4}+\frac{1}{4}e^{-2\Gamma_{\text{eff}}''t}\right)
\nonumber \\
+\rho_{\text{eff},3,3}(0)\left(\frac{3}{8}+\frac{1}{2}e^{-\frac{3}{2}\Gamma_{\text{eff}}''t}+\frac{1}{8}e^{-2\Gamma_{\text{eff}}''t}\right).
\end{eqnarray}
Since the steady state depends on the initial state, a limit cycle state does not exist.
Specifically, notice that steady state $\hat{\rho}_{\text{ss}}$ is equal to $|2\rangle \langle2|$ if and only if $\hat{\rho}(t=0)=|2\rangle\langle2|$.  
The steady-state solution depends on the phases of the coupling lasers.

\subsection{Case (ii): $\Delta_B=0$, $\phi_{\pm 1}=0$,  and $\Omega'$ finite}
The effective 3-level master equation reads
\begin{eqnarray}
    \dot{\rho}_{\text{eff},1,1}(t)
    = \Gamma_{\text{eff}}''
    \left[ -\rho_{\text{eff},1,1}(t)-\frac{1}{4}\rho_{\text{eff},1,3}(t)-\frac{1}{4}\rho_{\text{eff},3,1}(t)+\frac{1}{2}\rho_{\text{eff},3,3}(t)\right]
    -
    \Gamma_{\text{eff}}' \rho_{\text{eff},1,1}(t),
\end{eqnarray}
\begin{eqnarray}
    \dot{\rho}_{\text{eff},1,2}(t)
    = \Gamma_{\text{eff}}''
    \left[ -\frac{3}{4}\rho_{\text{eff},1,2}(t)-\frac{3}{4}\rho_{\text{eff},3,2}(t)\right]
    -
    \Gamma_{\text{eff}}' \rho_{\text{eff},1,2}(t),
\end{eqnarray}
\begin{eqnarray}
    \dot{\rho}_{\text{eff},1,3}(t)
    = \Gamma_{\text{eff}}''
    \left[ -\frac{3}{4}\rho_{\text{eff},1,1}(t)-\frac{3}{2}\rho_{\text{eff},1,3}(t)-\frac{3}{4}\rho_{\text{eff},3,3}(t)\right]
    -
    2\Gamma_{\text{eff}}' \rho_{\text{eff},1,3}(t),
\end{eqnarray}
\begin{eqnarray}
    \dot{\rho}_{\text{eff},2,2}(t)
    = \Gamma_{\text{eff}}''
    \left[ \frac{1}{2}\rho_{\text{eff},1,1}(t)+\frac{1}{2}\rho_{\text{eff},1,3}(t)+\frac{1}{2}\rho_{\text{eff},3,1}(t)+\frac{1}{2}\rho_{\text{eff},3,3}(t)\right]
    +
    \Gamma_{\text{eff}}' \left[ \rho_{\text{eff},1,1}(t)+\rho_{\text{eff},3,3}(t)
    \right],
\end{eqnarray}
\begin{eqnarray}
    \dot{\rho}_{\text{eff},2,3}(t)
    = \Gamma_{\text{eff}}''
    \left[ -\frac{3}{4}\rho_{\text{eff},2,1}(t)-\frac{3}{4}\rho_{\text{eff},2,3}(t)\right]
    -
    \Gamma_{\text{eff}}' \rho_{\text{eff},2,3}(t),
\end{eqnarray}
and
\begin{eqnarray}
    \dot{\rho}_{\text{eff},3,3}(t)
    = \Gamma_{\text{eff}}''
    \left[ \frac{1}{2}\rho_{\text{eff},1,1}(t)-\frac{1}{4}\rho_{\text{eff},1,3}(t)-\frac{1}{4}\rho_{\text{eff},3,1}(t)-\rho_{\text{eff},3,3}(t)\right]
    -
    \Gamma_{\text{eff}}' \rho_{\text{eff},3,3}(t),
\end{eqnarray}
where $\Gamma_{\text{eff}}''$ is given in Eq.~(\ref{eq_gamma_eff}) and
\begin{equation}
    \Gamma_{\text{eff}}'=\frac{|\Omega'|^2}{2\Gamma_{\text{aux}}'}.
\end{equation}
The analytical solution for  arbitrary $\underline{\rho}_{\text{eff}}(0)$ 
is quite long. Instead of reporting the full result, we only write out $\rho_{\text{eff},2,2}(t)$ and $\rho_{\text{eff},1,3}(t)$:
\begin{eqnarray}
\rho_{\text{eff},2,2}(t)=1-\frac{1}{2(\Gamma_{1})^2}e^{(-\Gamma_{\text{eff}}''-\frac{3}{2}\Gamma_{\text{eff}}'+\frac{1}{2}\Gamma_{1})t}\left[-2\Gamma_{\text{eff}}''\Gamma_{1}\Re\left(\rho_{\text{eff},1,3}(0)\right)+(\Gamma_{2,+})^2\left(\rho_{\text{eff},1,1}(0)+\rho_{\text{eff},3,3}(0)\right)\right]
\nonumber \\
+\frac{1}{2(\Gamma_{1})^2}e^{(-\Gamma_{\text{eff}}''-\frac{3}{2}\Gamma_{\text{eff}}'-\frac{1}{2}\Gamma_{1})t}\left[\left(-(\Gamma_{1})^2+\Gamma_{\text{eff}}''\Gamma_{1}+\Gamma_{\text{eff}}'\Gamma_{1}\right)\left(\rho_{\text{eff},1,1}(0)+\rho_{\text{eff},3,3}(0)\right)-2\Gamma_{\text{eff}}''\Gamma_{1}\Re\left(\rho_{\text{eff},1,3}(0)\right)\right]
\end{eqnarray}
and
\begin{eqnarray}
\rho_{\text{eff},1,3}(t)=-\frac{1}{4(\Gamma_{1})^2}e^{(-\Gamma_{\text{eff}}''-\frac{3}{2}\Gamma_{\text{eff}}'+\frac{1}{2}\Gamma_{1})t}\left[3\Gamma_{\text{eff}}''\Gamma_{1}\left(\rho_{\text{eff},1,1}(0)+\rho_{\text{eff},3,3}(0)\right)-2(\Gamma_{2,-})^2\Re\left(\rho_{\text{eff},1,3}(0)\right)\right]
\nonumber \\
+\frac{1}{4(\Gamma_{1})^2}e^{(-\Gamma_{\text{eff}}-\frac{3}{2}\Gamma_{\text{eff}}'-\frac{1}{2}\Gamma_{1})t}\left[3\Gamma_{\text{eff}}''\Gamma_{1}\left(\rho_{\text{eff},1,1}(0)+\rho_{3,3}(0)\right)+2\Gamma_{2,+}^2\Re(\rho_{\text{eff},1,3}(0))\right]+i e^{(-\frac{3}{2}\Gamma_{\text{eff}}''-2\Gamma_{\text{eff}}')t}\Im\left(\rho_{\text{eff},1,3}(0)\right),\nonumber \\
\end{eqnarray}
\end{widetext}
where
\begin{equation}
    \Gamma_{1}=\sqrt{4(\Gamma_{\text{eff}}'')^2+2\Gamma_{\text{eff}}''\Gamma_{\text{eff}}'+(\Gamma_{\text{eff}}')^2}
\end{equation}
and
\begin{equation}
    \Gamma_{2,\pm}=\sqrt{4(\Gamma_{\text{eff}}'')^2+\Gamma_{\text{eff}}'(\Gamma_{\text{eff}}'\pm\Gamma_{1})+\Gamma_{\text{eff}}''(2\Gamma_{\text{eff}}'\pm\Gamma_{1})}.
\end{equation}
The steady state limits of $\rho_{\text{eff},2,2}(t)$ and $\rho_{\text{eff},1,3}(t)$ are $1$ and $0$, respectively.
The full solutions prove that the limit cycle state is $|2\rangle \langle 2|$. We know from our  numerical investigations that this is also the limit cycle state for finite $\phi_{\pm 1}$. Intuitively, this can be understood by realizing that the decay beam forces the reference system, which is characterized by  $\Omega_{0}=0$, into the limit cycle
 state $\hat{\rho}_{\text{ss}}=|2\rangle\langle2|$, regardless of the laser phases. The reason is that the decay laser couples to states that are not coupled by other lasers. 

\subsection{Case (iii): $\Delta_B$ finite, $\Omega_{\pm 1}=0$, and $\Omega'$ finite}
The effective 3-level master equation reads
\begin{eqnarray}
    \dot{\rho}_{\text{eff},1,1}(t)= - \Gamma_{\text{eff}}' \rho_{\text{eff},1,1}(t) ,
\end{eqnarray}
\begin{eqnarray}
    \dot{\rho}_{\text{eff},1,2}(t)= - \Gamma_{\text{eff}}' \rho_{\text{eff},1,2}(t) -
    i \Delta_{\text{eff}}' \rho_{\text{eff},1,2}(t),
\end{eqnarray}
\begin{eqnarray}
    \dot{\rho}_{\text{eff},1,3}(t)= - 2\Gamma_{\text{eff}}' \rho_{\text{eff},1,3}(t) -
    2i \Delta_{\text{eff}}' \rho_{\text{eff},1,3}(t),
\end{eqnarray}
\begin{eqnarray}
    \dot{\rho}_{\text{eff},2,2}(t)= \Gamma_{\text{eff}}' \left[ \rho_{\text{eff},1,1}(t) +\rho_{\text{eff},3,3}(t) \right],
\end{eqnarray}
\begin{eqnarray}
    \dot{\rho}_{\text{eff},2,3}(t)= - \Gamma_{\text{eff}}' \rho_{\text{eff},2,3}(t) -
    i \Delta_{\text{eff}}' \rho_{\text{eff},2,3}(t),
\end{eqnarray}
and
\begin{eqnarray}
    \dot{\rho}_{\text{eff},3,3}(t)= - \Gamma_{\text{eff}}' \rho_{\text{eff},3,3}(t) ,
\end{eqnarray}
where
\begin{equation}
\label{eq_case3_a}
    \Gamma_{\text{eff}}'=\frac{\Gamma_{\text{aux}}'}{2}\frac{|\Omega'/2|^2}{(\Delta_{B}-\Delta_{B}')^2+(\Gamma_{\text{aux}}'/2)^2}
\end{equation}
and
\begin{equation}
\label{eq_case3_b}
    \Delta_{\text{eff}}'=-\Delta_B-\frac{(\Delta_{B}-\Delta_{B}') |\Omega'|^2}{(\Gamma_{\text{aux}}')^2+4(\Delta_{B}-\Delta_{B}')^2}.
\end{equation}
For $B=0$, Eqs.~(\ref{eq_case3_a}) and (\ref{eq_case3_b}) reduce to $\Gamma_{\text{eff}}'=\frac{|\Omega'|^2}{2\Gamma_{\text{aux}}'}$ and $\Delta_{\text{eff}}'=0$, respectively.
Since the differential equations for the matrix elements $\rho_{\text{eff},j.k}(t)$ are decoupled from each other, 
the solutions for an  arbitrary initial state $\underline{\rho}_{\text{eff}}(0)$ can be obtained readily,
\begin{align}
\rho_{\text{eff},1,1}(t)=\rho_{\text{eff},1,1}(0)e^{-\Gamma_{\text{eff}}'t},
\\
\rho_{\text{eff},1,2}(t)=\rho_{\text{eff},1,2}(0)e^{-(\Gamma_{\text{eff}}'+i \Delta_{\text{eff}}')t},
\\
\rho_{\text{eff},1,3}(t)=\rho_{\text{eff},1,3}(0)e^{-2(\Gamma_{\text{eff}}'+i \Delta_{\text{eff}}')t},
\\
\rho_{\text{eff},2,2}(t)=1- \left[ \rho_{\text{eff},1,1}(0)+\rho_{3,3}(0) \right] e^{-\Gamma_{\text{eff}}'t},
\\
\rho_{\text{eff},2,3}(t)=\rho_{\text{eff},2,3}(0)e^{-(\Gamma_{\text{eff}}'+i \Delta_{\text{eff}}')t},
\\
\rho_{\text{eff},3,3}(t)=\rho_{\text{eff},3,3}(0)e^{-\Gamma_{\text{eff}}'t}.
\end{align}
 Notice that the steady-state density operator
$ \hat{\rho}_{\text{ss}}$ is equal to $|2\rangle \langle2|$.
It can be shown that this is also the limit cycle state for finite $\Omega_{\pm1}$.


\begin{thebibliography}{98}
\bibitem{footnote1}
We use $S$ to denote the spin of a generic spin system and $F$ to denote the spin of a specific atomic hyperfine manifold.
\end{thebibliography}
\end{document}